\documentclass[letterpaper,twocolumn,american,floatfix,amsmath,amssymb,reprint,nofootinbib,pre,superscriptaddress]{revtex4-1}
\usepackage[T1]{fontenc}
\usepackage[latin9]{inputenc}
\usepackage{color}
\usepackage{babel}
\usepackage{amsmath}
\usepackage{amssymb}
\usepackage{graphicx}
\usepackage{wasysym}
\usepackage{esint}
\usepackage[unicode=true,pdfusetitle,
 bookmarks=false,
 breaklinks=true,pdfborder={0 0 0},backref=false,colorlinks=true]
 {hyperref}
\hypersetup{
 linkcolor=blue,citecolor=blue,urlcolor=blue}
\usepackage{breakurl}

\makeatletter

\usepackage{babel}

\IfFileExists{lmodern.sty}{\usepackage{lmodern}}{}

\makeatother

\begin{document}

\title{Temporal disorder in discontinuous non-equilibrium phase transitions:
general results}

\author{C. E. Fiore}

\affiliation{Instituto de Física, Universidade de São Paulo, Caixa Postal 66318
São Paulo, São Paulo 05315-970, Brazil.}

\author{M. M. de Oliveira}

\affiliation{Departamento de Física e Matemática, CAP, Universidade Federal de
São João del Rei, Ouro Branco, Minas Gerais 36420-000, Brazil.}

\author{José A. Hoyos}

\affiliation{Instituto de Física de São Carlos, Universidade de São Paulo, C.P.
369, São Carlos, São Paulo 13560-970, Brazil.}

\date{\today}
\begin{abstract}
We develop a general theory for discontinuous non-equilibrium phase
transitions into an absorbing state in the presence of temporal disorder.
We focus in two paradigmatic models for discontinuous transitions:
the quadratic contact process (in which activation is only spread
when two nearest-neighbor sites are both active) and the contact process
with long-range interactions. Using simple stability arguments (supported
by Monte Carlo simulations), we show that temporal disorder does not
destroy the discontinuous transition in the former model. For the
latter one, the first-order transition is turned into a continuous
one only in the strong-disorder limit, with critical behavior belonging
to the infinite-noise universality class of the contact process model.
Finally, we have found that rare temporal fluctuations dramatically
changes the behavior of metastable phase turning it into a temporal
Griffiths inactive phase characterized by an exponentially large decay
time.
\end{abstract}
\maketitle

\section{Introduction}

Non-equilibrium phase transitions have constituted a rich and lively
topic of research for many years. They occur in a wide variety of
models in ecology~\cite{kussell-vucelja-rpp14}, epidemic spreading~\cite{pastor-etal-rmp15},
sociophysics~\cite{sen-chakrabarti-book13}, catalytic reactions~\cite{ziff-gulari-barshad-prl86},
depinning interface growth~\cite{tang-leschhorn-pra92,buldyrev-etal-pra92},
turbulent flow~\cite{pomeau-pd86}, among other fields~\cite{marro-dickman-book,henkel-etal-book08,odor-rmp-04}.

Since disorder due to spatial or temporal inhomogeneities is almost
an unavoidable ingredient in many real systems, it is then desirable
to understand their effects on these phase transitions. For continuous
phase transitions, it was earlier recognized that spatial and temporal
disorder changes the critical behavior whenever the generalized Harris
criterion is violated~\cite{kinzel-zpb85,noest-prl86}: quenched
spatial disorder is relevant whenever $d\nu_{\perp}>2$ is violated
while temporal disorder is relevant when $\nu_{\parallel}=z\nu_{\perp}>2$
is violated; with $\nu_{\perp},$ $\nu_{\parallel}$ and $z$ being
critical exponents of the clean phase transition and $d$ being the
number of spatial dimensions. Since the critical exponents of the
directed percolation universality class violate the Harris criterion,
it was then argued that this was the reason why it was never seen
in experiments~\cite{hinrichsen-bjp00} (see however Ref.~\onlinecite{takeuchi-etal-prl07}).

Later, it was shown that spatial disorder yields a critical behavior
in the exotic universality class of infinite-randomness type surrounding
accompanied by a Griffiths effects in the inactive phase~\cite{hooyberghs-prl,vojta-dickison-nonequ,vojta-lee-prl06,hoyos-pre08,oliveira-ferreira-jsm08,vojta-etal-pre09}.
More recently, it was shown that temporal disorder yields to analogous
effects, namely, an exotic infinite-noise universality class accompanied
by a temporal Griffiths active phase~\cite{vazquez-etal-prl11,vojta-hoyos-epl15,barghathi-etal-pre16,solano-etal-pre16}. 

The effects of disorder in discontinuous non-equilibrium phase transitions
are much less understood. It was initially shown that quenched spatial
disorder can turn a discontinuous transition into a continuous one~\cite{hoenicke-figueiredo-pre00}
and later, it was argued that it actually prohibits phase coexistence
and discontinuous transitions in $d\le2$~\cite{martin-etal-pre14}.
In the case of temporal disorder, however, a recent numerical study
indicates that first-order phase transitions can happen in low-dimensional
systems~\cite{oliveira-fiore-pre16}.

In this work, we develop a general theory for discontinuous non-equilibrium
phase transition in the presence of temporal disorder. Analysis of
two paradigmatic models in mean-field level is sufficient to draw
quantitative accurate predictions which we confirm in $d=1$ and $2$
via Monte Carlo simulations. Our main result is that temporal disorder
does not forbid first-order phase transitions. In addition, it can
also turn a discontinuous transition into a continuous one when disorder
is sufficiently strong. Furthermore, we find an interesting novel
phenomena: temporal disorder turns the clean metastable active phase
into a temporal Griffiths inactive phase characterized by extremely
large decay times. 

The remainder of this article, we define the studied models in Sec.~\ref{sec:The-models},
develop our main theory in Sec.~\ref{sec:The-mean-field-approach}
where a mean-field analysis is performed. In Sec.~\ref{sec:Monte-Carlo-simulations}
we provide Monte Carlo simulations confirming our theory and leave
concluding remarks to Sec.~\ref{sec:Decaying-time}.

\section{The models\label{sec:The-models}}

The usual contact process (CP) model~\cite{harris-an74,marro-dickman-book}
is defined on a $d$-dimensional lattice in which each site is either
active ($A$) or inactive ($I$). The corresponding dynamics has the
following processes: (i) a spontaneous inactivation and (ii) an autocatalytic
activation via nearest-neighbor contact. In the former, a single active
site spontaneously decays to the inactive state with rate $\mu$.
In the latter, an active site turns an inactive nearest-neighbor site
into an active one with rate $\lambda$. Schematically, $A\stackrel{\mu}{\rightarrow}I$
and $A+I\stackrel{\lambda}{\rightarrow}2A$, respectively.

In this work, we study a particular case of the second Schlögl model~\cite{Schlogl-zfp72},
known as the quadratic contact process (QCP) model and a version of
the CP model with long-range interactions known as the $\sigma$CP
model~\cite{ginelli-etal-pre05}. They are identical to the CP model
except for the activity spreading dynamics. In the QCP model, activity
is spread via the contact with two active nearest-neighbor sites:
$2A+I\stackrel{\lambda}{\rightarrow}3A$. In the $\sigma$CP model,
the activation rate depends on the length $\ell$ of the continuous
string of inactive sites between two active ones, i.e., $\lambda\rightarrow\lambda_{\ell}=\lambda\left(1+a\ell^{-\sigma}\right)$,
where $a\geq0$ and $\sigma>0$ are constants controlling the long-range
``interaction'' (with $a=0$ recovering the CP model). Schematically,
the reaction is $A+I^{\ell}\stackrel{\lambda_{\ell}}{\rightarrow}2A+I^{\ell-1}$,
where $I^{\ell}$ denotes the continuous string of $\ell$ inactive
sites.

For simplicity, we set $\mu+\lambda=1$ and only deals with $\lambda\in[0,1]$.

Noise fluctuations (temporal disorder) are introduced in these models
by considering $\lambda$ as a random time-dependent variable. For
concreteness, we divide the system time evolution in time intervals
of equal duration $\Delta t$ within which $\lambda$ is constant,
i.e., over the $i$-th time interval the activity spreading rate equals
$\lambda=\lambda_{i}$, with $\lambda_{i}$ being an independent random
variable drawn from a binary probability density distribution 
\begin{equation}
P(\lambda)=p\delta(\lambda-\lambda_{-})+\left(1-p\right)\delta(\lambda-\lambda_{+}),\label{eq:binary-dist}
\end{equation}
 with $\lambda_{+}>\lambda_{-}$. For later convenience, we rewrite
$\lambda_{\pm}$ in terms of the average $\overline{\lambda}=p\lambda_{-}+\left(1-p\right)\lambda_{+}$
and $\delta\lambda=\lambda_{+}-\lambda_{-}$ (which represents the
disorder strength), namely, $\lambda_{+}=\overline{\lambda}+p\delta\lambda$
and $\lambda_{-}=\overline{\lambda}-\left(1-p\right)\delta\lambda$.
We report that we have also considered box-like distributions and
have found no qualitative difference.

\section{The mean-field approach\label{sec:The-mean-field-approach}}

In this Section, we present our mean-field approach for the effects
of temporal disorder on the first-order non-equilibrium phase transitions
to an absorbing state.

\subsection{The clean system}

We start by reviewing some key aspects of the clean phase transition
and later consider the effects of temporal disorder.

\subsubsection{Mean-field approach for the clean QCP model\label{sub:Mean-field-QCP}}

Let us start with the QCP model at the level of one-site mean-field
theory. The density of active sites $\rho$ obeys the following logistic
equation 
\begin{equation}
\frac{{\rm d}\rho}{{\rm d}t}=-(1-\lambda)\rho+\lambda\rho^{2}(1-\rho),\label{eq:dr-dt-QCP}
\end{equation}
 where the first term on the RHS accounts for the spontaneous inactivation
processes, whereas the second one corresponds to the activity spreading.

There are three steady-state (time-independent) solutions $\rho_{\infty}$
for Eq.~\eqref{eq:dr-dt-QCP}: 
\begin{equation}
\rho_{\infty}^{(I)}=0,\mbox{ }\rho_{\infty}^{(S)}=\frac{1}{2}+\alpha,\mbox{ and }\rho_{\infty}^{(U)}=\frac{1}{2}-\alpha,\label{eq:r-infinity-QCP}
\end{equation}
 with $\alpha=\sqrt{\frac{5}{4}-\frac{1}{\lambda}}$. A phase transition
occurs at $\lambda=\lambda_{c}=\frac{4}{5}$ above which $\rho_{\infty}^{(S)}$
and $\rho_{\infty}^{(U)}$ exist ($\alpha\in\mathbb{R}$). As $\rho_{\infty}^{(S,U)}\rightarrow\frac{1}{2}$
when $\lambda\rightarrow\lambda_{c}^{+}$, notice the transition is
discontinuous with the order parameter being $\rho_{c}=\frac{1}{2}$
at the transition. In order to better understand the phases surrounding
the transition point, we study the stability of the steady-state solutions
$\rho_{\infty}$ by linearizing Eq.~\eqref{eq:dr-dt-QCP}. It is
found that $\rho_{\infty}^{(I)}$ is a stable solution for $0\leq\lambda<\lambda^{*}=1$
with small deviations from it ($r=\rho-\rho_{\infty}$) vanishing
exponentially $r\sim e^{-\left(1-\lambda\right)t}$ for large $t$.
Likewise, $\rho_{\infty}^{(S)}$ is a stable solution (for $\lambda>\lambda_{c}$)
with small deviations vanishing as $\left|r\right|\sim e^{-\left[\left(\frac{5}{2}+\alpha\right)\lambda-2\right]t}$
for large $t$. Finally, the $\rho_{\infty}^{(U)}$ is an unstable
solution (for $\lambda>\lambda_{c}$) in which deviations grow as
$\left|r\right|\sim e^{\left[2-\left(\frac{5}{2}-\alpha\right)\lambda\right]t}$
for small $t$. At the transition point $\lambda=\lambda_{c}$, the
solutions $\rho_{\infty}^{(S,U)}$ degenerate and become a saddle
point. In this case when $\rho>\rho_{c}=\rho_{\infty}^{(S,U)}$, the
deviations vanish algebraically as $r\sim t^{-1}$ for large $t$,
otherwise when $\rho<\rho_{c}$, they increase as $r\approx-\left|r_{0}\right|(1+\left|r_{0}\right|\rho_{c}\lambda_{c}t)$
for small $t$.

We call attention to the fact that for $\lambda_{c}\leq\lambda<\lambda^{*}$
there are two stable solutions $\rho_{\infty}^{(I,S)}$ being one
of them corresponding to the inactive absorbing state. As we show
latter, this bistability is an important feature for understanding
the temporal disorder effects. For this reason, we refer to this region
of the active phase as metastable phase.

\begin{figure}
\begin{centering}
\includegraphics[clip,width=0.8\columnwidth]{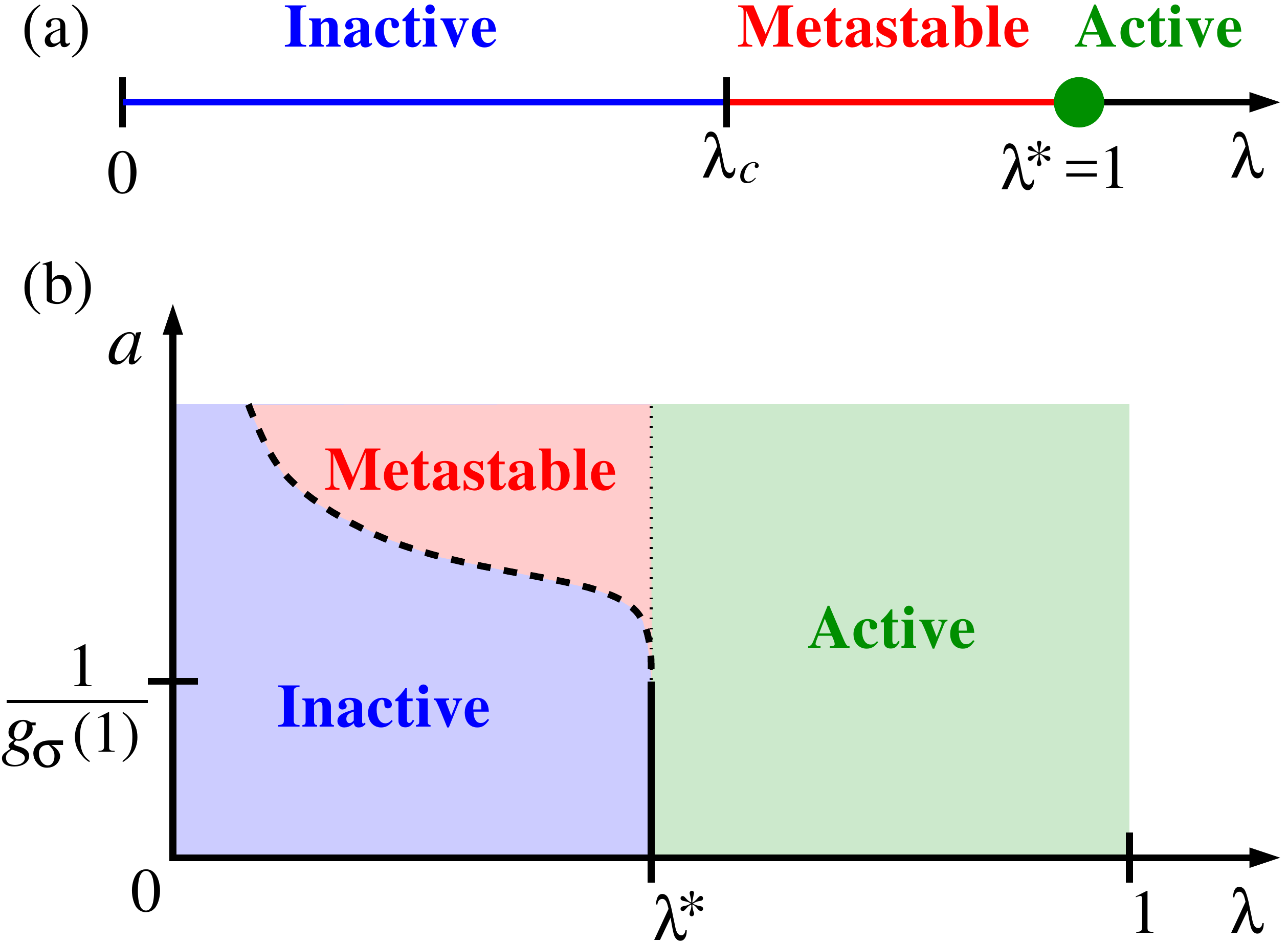} 
\par\end{centering}

\caption{Mean-field phase diagram of the clean (a) QCP and (b) $\sigma$CP
models (see main text). The dashed line denotes a first-order phase
transition and the solid line denotes a second order one belonging
to the directed percolation universality class. The dotted line denotes
the end of the bistability in the active phase. For the QCP model,
$\lambda_{c}=\frac{4}{5}$. For the $\sigma$CP model, $\lambda^{*}=\frac{1}{2}$.
\label{fig:clean-PD}}
\end{figure}

The mean-field phase diagram of the QCP model is shown in Fig.\ \hyperref[fig:clean-PD]{\ref{fig:clean-PD}(a)}.
For $0\leq\lambda<\lambda_{c}$, the system is in the inactive phase
in which any activity becomes extinct as $t\rightarrow\infty$ with
$\rho\rightarrow\rho_{\infty}^{(I)}$. For $\lambda_{c}<\lambda<1$,
the system is in the metastable phase in which activity persists ($\rho\rightarrow\rho_{\infty}^{(S)}$)
indefinitely if the initial density $\rho(0)\equiv\rho_{0}$ is greater
than $\rho_{\infty}^{(U)}$, otherwise the system evolves towards
the absorbing state. The transition at $\lambda=\lambda_{c}$ between
the inactive and the metastable phase is discontinuous. Finally, at
$\lambda=\lambda^{*}=1$ the system is in the usual active phase.

It worth noting that Eq.~\eqref{eq:dr-dt-QCP} can be fully integrated,
yielding 
\begin{align}
\ln\left(\frac{\rho}{\rho_{0}}\right)-\left(\frac{\rho_{\infty}^{(S)}}{2\alpha}\right)\ln\left(\frac{\rho-\rho_{\infty}^{(U)}}{\rho_{0}-\rho_{\infty}^{(U)}}\right)\nonumber \\
+\left(\frac{\rho_{\infty}^{(U)}}{2\alpha}\right)\ln\left(\frac{\rho-\rho_{\infty}^{(S)}}{\rho_{0}-\rho_{\infty}^{(S)}}\right) & =-\left(1-\lambda\right)t.\label{eq:rho-t-exact-QCP}
\end{align}
 From this solution, all previous conclusions follow straightforwardly.
Evidently, at the transition point $\lambda=\lambda_{c}$, a direct
integration of the resulting logistic equation $\frac{{\rm d}\rho}{{\rm d}t}=-\lambda_{c}\rho(\rho-\rho_{c})^{2}$
yields to 
\begin{equation}
\ln\left(\frac{\rho_{0}\left(\rho-\rho_{c}\right)}{\rho\left(\rho_{0}-\rho_{c}\right)}\right)+\frac{\rho_{c}\left(\rho_{0}-\rho\right)}{\left(\rho-\rho_{c}\right)\left(\rho_{0}-\rho_{c}\right)}=\rho_{c}^{2}\lambda_{c}t.
\end{equation}

\paragraph{Decay time towards the absorbing state close to the transition: general
results}

\begin{figure}[t]
\begin{centering}
\includegraphics[clip,width=0.8\columnwidth]{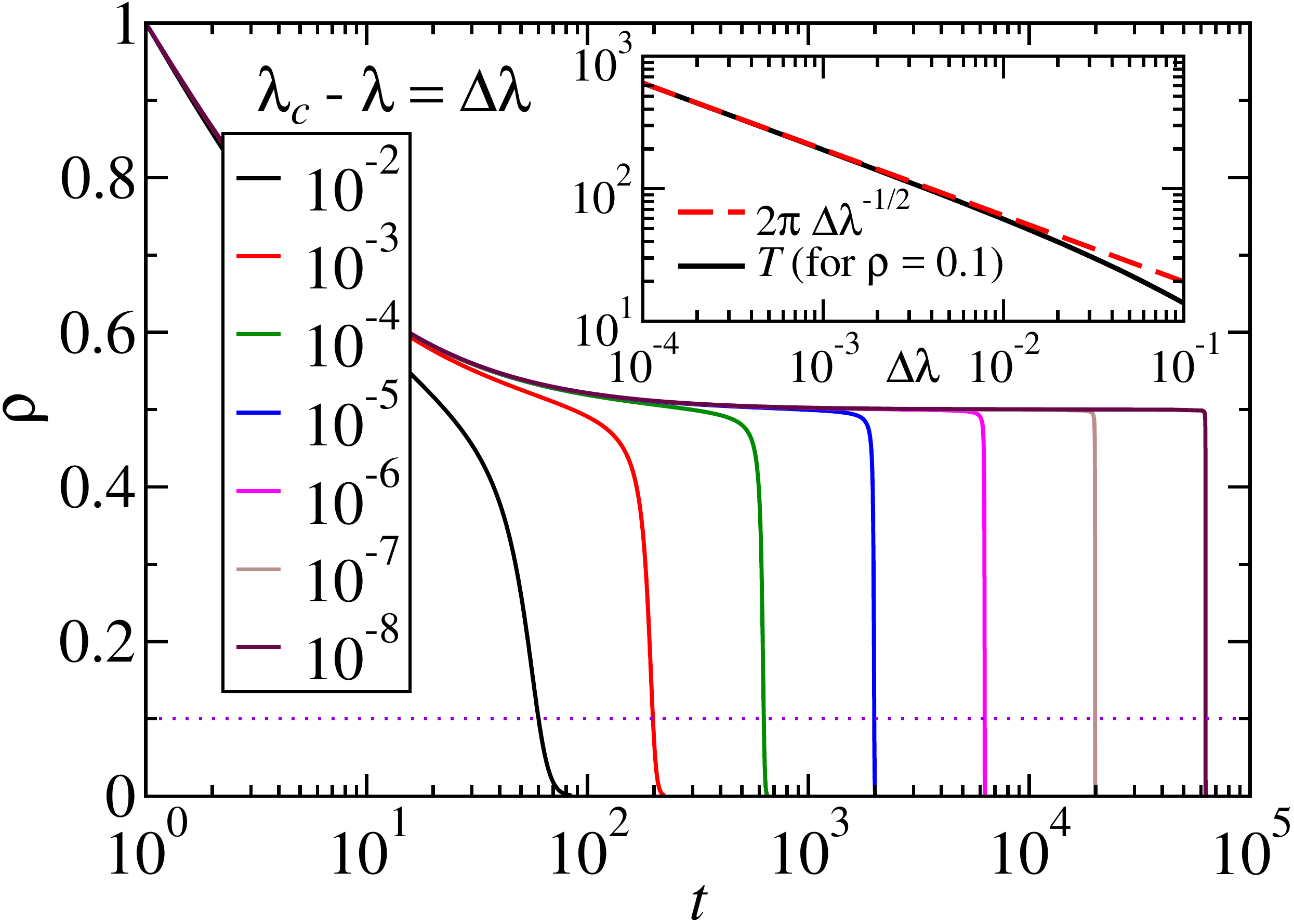}
\par\end{centering}

\caption{The mean-field density $\rho$ as a function of time $t$ for the
QCP model for various activation rates $\lambda$ in the inactive
phase $\lambda<\lambda_{c}$. The inset shows the time $T$ when $\rho=0.1$
(dotted line of the main panel). The dashed line is the analytical
result Eq.~\eqref{eq:T}.\label{fig:time-T}}
\end{figure}

An important quantity for our analysis is the time $T$ necessary
for the system to decay into the absorbing state when it is in the
inactive phase but very close to the transition, i.e., when $\lambda=\lambda_{c}-\ell$,
with $0<\ell\ll\lambda_{c}$ (see Fig.~\ref{fig:time-T}). A intuitive
definition for $T$ would be the following: starting from $\rho_{0}=1,$
the decay time $T$ is such that$\rho(T)\ll\rho_{c}$. Although this
can be easily accomplished, we adopt another (and more elegant) one:
We define $T$ as the time interval for the evolution from $\rho_{0}=\rho_{c}+\epsilon$
to $\rho(T)=\rho_{c}-\epsilon$, with $0<\epsilon\ll\rho_{c}$. Since
we now have to small parameters $\frac{\ell}{\lambda_{c}}$ and $\frac{\epsilon}{\rho_{c}}$,
we now need to specify which one is smaller. Since we wish to connect
with the first definition, we then require that $\frac{\ell}{\lambda_{c}}\lll\frac{\epsilon}{\rho_{c}}$.
Inspections o the resulting logistic equation show that $\ell\rho_{c}^{2}\ll\epsilon^{2}\lambda_{c}$
is sufficient. 

We are now in position of computing $T$. This task can accomplish
in more general grounds (applicable to other models) by considering
a logistic equation of type 
\begin{equation}
\frac{{\rm d}\rho}{{\rm d}t}=\rho\left(\lambda f(\rho)-1\right).\label{eq:logistic-eq}
\end{equation}
 (The choice $f=1+\rho-\rho^{2}$ recovers the QCP model.) The discontinuous
transition point $\lambda_{c}$ and density $\rho_{c}$ are obtained
from $\lambda_{c}f(\rho_{c})=1$ and $f^{\prime}(\rho_{c})=0$. Defining
$\rho(t)=\rho_{c}-r(t)$, we study the time $T$ required for $r(t)$
evolving from $-\epsilon$ to $\epsilon$. Expanding the logistic
equation \eqref{eq:logistic-eq} for $\left|r\right|$ and $\ell$
(and noticing that $f^{\prime\prime}(\rho_{c})<0$), then 
\begin{equation}
\frac{\text{d}r}{r^{2}+R^{2}}\approx\frac{\pi}{RT}\text{d}t,\label{eq:r-t-approximate}
\end{equation}
 where $R=\sqrt{\frac{2\ell}{\left|f^{\prime\prime}(\rho_{c})\right|}}f(\rho_{c})$
and 
\begin{equation}
T=\frac{2\pi}{\sqrt{2\ell\left|f^{\prime\prime}(\rho_{c})\right|}\rho_{c}}=t_{0}\left(\lambda_{c}-\lambda\right)^{-\phi}.\label{eq:T}
\end{equation}
 The time scale $T$ is exactly the decay time obtained by integrating
Eq.~\eqref{eq:r-t-approximate} from $-\epsilon$ to $\epsilon$
and taking the limit $\epsilon\gg R$. We finally conclude that, in
the mean-field approximation, $T$ diverges with exponent $\phi=\frac{1}{2}$.
(For the QCP model, the microscopic time scale is $t_{0}=2\pi$, see
also Fig.~\eqref{fig:time-T}).

\subsubsection{Mean-field approach for the clean $\sigma$CP model}

In this case, at the level of one-site mean-field theory, the density
of active sites $\rho$ is obtained from

\begin{align}
\frac{{\rm d}\rho}{{\rm d}t} & =-(1-\lambda)\rho+\lambda\rho^{2}\sum_{{\rm \ell=1}}^{\infty}(1+a\ell^{-\sigma})(1-\rho)^{\ell}\nonumber \\
 & =(2\lambda-1)\rho-\lambda\rho^{2}\left[1-ag_{\sigma}(1-\rho)\right],\label{eq:dr-dt-sCP}
\end{align}
 where $g_{\nu}(z)=\sum_{\ell=1}^{\infty}\frac{z^{\ell}}{\ell^{\nu}}=\frac{1}{\Gamma(\nu)}\int_{0}^{\infty}\frac{x^{\nu-1}}{z^{-1}e^{x}-1}{\rm d}x$
is the Polylogarithm function which, for $0\leq z<1$ and $\nu>0$,
becomes the familiar Bose-Einstein function. Notice that when $ag_{\sigma}\left(1-\rho\right)\geq1$,
the nonlinear term $\propto\rho^{2}$ changes sign and a new behavior
is expected, otherwise the same physics of the usual CP model is recovered.
Finally, notice that Eq.~\eqref{eq:dr-dt-sCP} is of the type \eqref{eq:logistic-eq}
with $f=2-\rho\left(1-ag_{\sigma}\left(1-\rho\right)\right)$.

\begin{figure}[t]
\centering{}\includegraphics[clip,width=0.8\columnwidth]{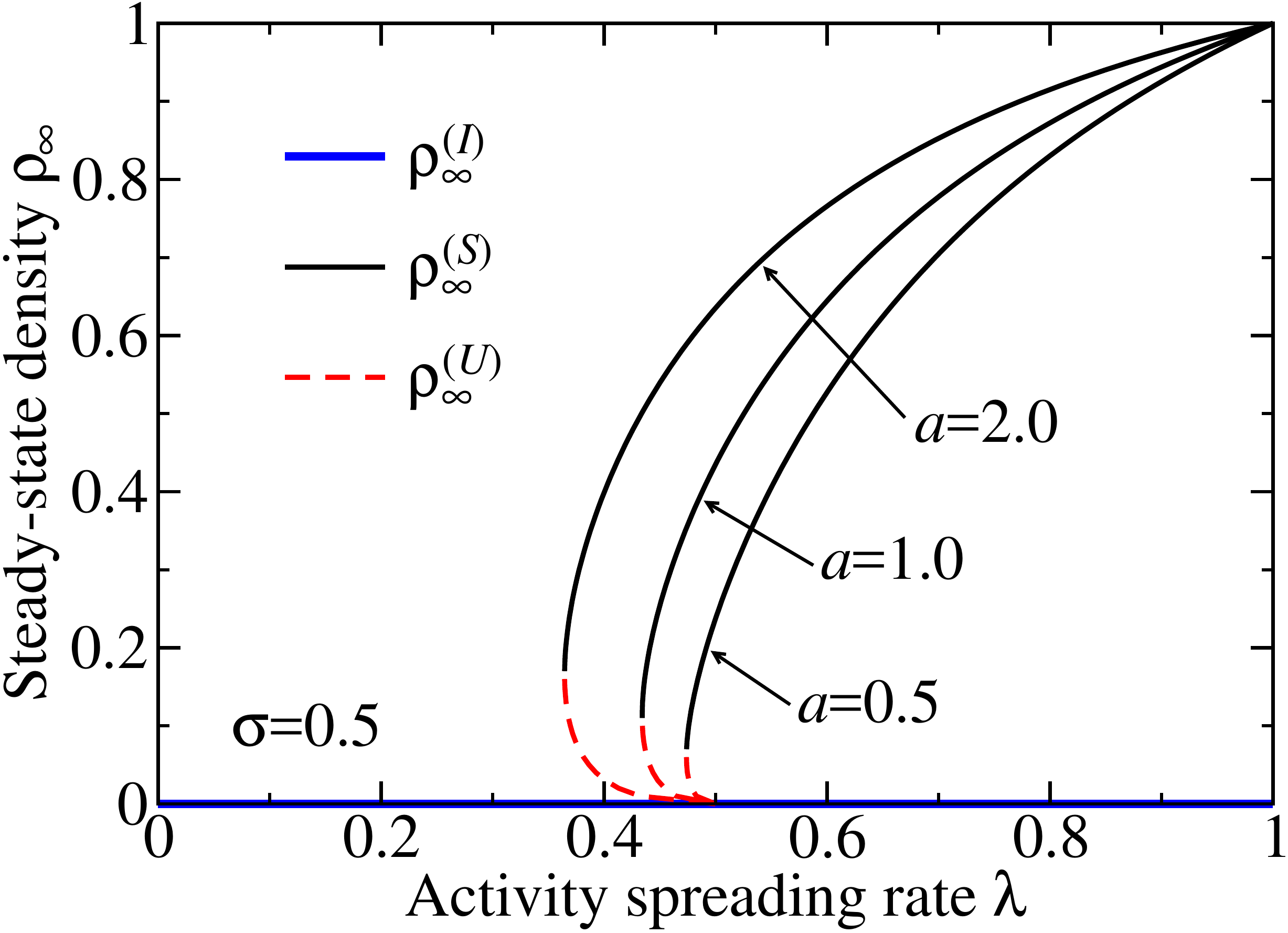}\\
 \includegraphics[clip,width=0.8\columnwidth]{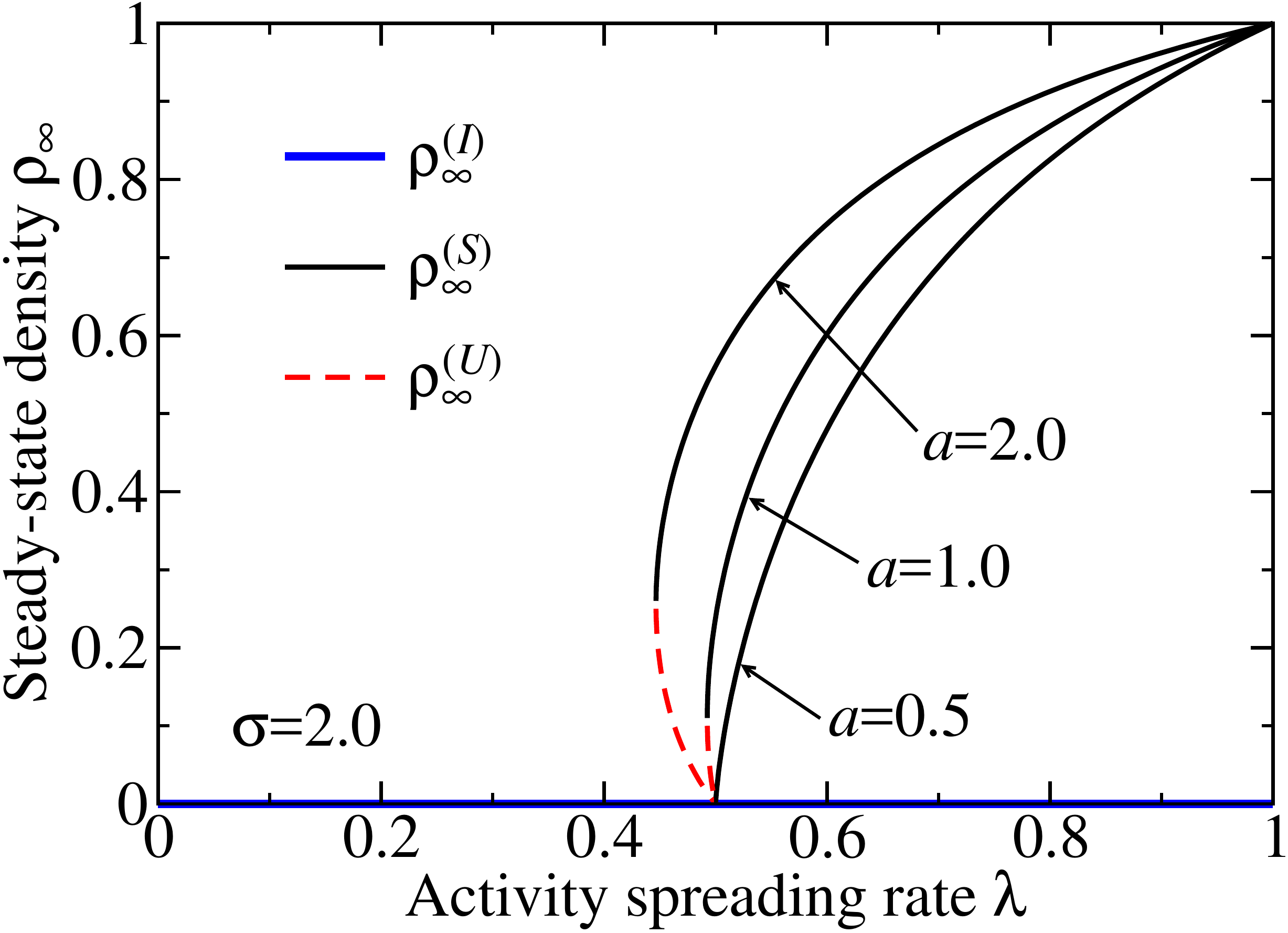}
\caption{The possible steady-state densities $\rho_{\infty}$ as a function
of the activity spreading rate $\lambda$ for the values of the exponent
$\sigma=0.5$ (top) and $\sigma=2.0$ (bottom panel) and various values
of the parameter $a$ as indicated.\label{fig:rho-lambda-scpmodel}}
\end{figure}

As in the QCP model, there is a trivial steady-state density $\rho_{\infty}^{(I)}=0$
representing the inactive absorbing state. It is stable for $\lambda\leq\lambda^{*}=\frac{1}{2}$
and unstable for $\lambda>\lambda^{*}$. Thus, $\lambda>\lambda^{*}$
delimits the usual active phase (without bistability). Nontrivial
steady-state densities are shown in Fig.~\ref{fig:rho-lambda-scpmodel}
for some values of $a$ and $\sigma$, which are the real solutions
of the equation $f(\rho_{\infty})=\lambda^{-1}$, namely 
\begin{equation}
\rho_{\infty}\left[1-ag_{\sigma}(1-\rho_{\infty})\right]=2-\lambda^{-1}.\label{eq:stead-state-sCP}
\end{equation}
 When $\lambda>\lambda^{*}$, Eq.~\eqref{eq:stead-state-sCP} has
only one stable solution $\rho_{\infty}^{(S)}$ corresponding to the
usual active phase as already anticipated. When $f^{\prime}\left(0\right)>0$
(or $ag_{\sigma}(1)>1$), Eq.~\eqref{eq:stead-state-sCP} has two
finite-density steady-state solutions: $\rho_{\infty}^{(S)}$ and
$\rho_{\infty}^{(U)}$ (with $\rho_{\infty}^{(S)}\geq\rho_{\infty}^{(U)}$)
which are stable and unstable, respectively. Thus, the region $\lambda_{c}<\lambda<\lambda^{*}$
corresponds to the metastable phase. At $\lambda=\lambda^{*}$, the
bistability of the active phase ends.

For $f^{\prime}(0)>0,$ it is clear the transition from the inactive
phase to a metastable phase at $\lambda=\lambda_{c}<\lambda^{*}$
is discontinuous. The order parameter $\rho_{c}$ at the discontinuous
transition is obtained from $f^{\prime}(\rho_{c})=0$, i.e., $a^{-1}=g_{\sigma}(1-\rho_{c})-\frac{\rho_{c}}{1-\rho_{c}}g_{\sigma-1}(1-\rho_{c})$.
The corresponding transition point is $\lambda_{c}=f^{-1}(\rho_{c})=[2-\rho_{c}(1-ag_{\sigma}(1-\rho_{c}))]^{-1}$
{[}see the dashed line in Fig.\ \hyperref[fig:clean-PD]{\ref{fig:clean-PD}(b)}{]}.
On the other hand if $ag_{\sigma}\left(1\right)\leq1$, the transition
from the inactive to the active phase is continuous at $\lambda=\lambda^{*}$
and belonging to the directed percolation universality class.

Finally, at the inactive phase but near the transition point to the
metastable phase, $\lambda=\lambda_{c}-\ell$, the time needed for
decay from an initial state such that $\rho(0)>\rho_{c}$ diverges
when $\ell\rightarrow0^{+}$ as $T=t_{0}\ell^{-\phi}$, with exponent
$\phi=\frac{1}{2}$ and constant $t_{0}=2\pi\left(\rho_{c}^{-1}-1\right)/\sqrt{2a\left[\left(2-\rho_{c}\right)g_{\sigma-1}-\rho_{c}g_{\sigma-2}\right]}$,
according to Eq.~\eqref{eq:T}.

\subsection{Overview of the temporal disorder effects\label{sub:Overview}}

Let us now discuss the effects of temporal disorder on the clean phase
diagram of the QCP and $\sigma$CP models (see Fig.~\ref{fig:clean-PD}).
For simplicity, we assume that $\lambda$ takes only two possible
distinct values with equal and independent probabilities {[}see Eq.~\eqref{eq:binary-dist}
for $p=1/2${]} along system time evolution. As will become clear,
although we base our quantitative conclusions on the mean-field analysis,
our conclusions are qualitatively applicable to any dimension provided
that it supports a discontinuous phase transition.

\subsubsection{Effects on the phases\label{sub:Effects-on-phases}}

Firstly, let us discuss the effects of temporal disorder on the nature
of the phases, i.e., let us discuss the case in which both $\lambda_{-}$
and $\lambda_{+}$ are in the same (clean) phase.

When $0\leq\lambda_{\pm}<\lambda_{c}$, the system inevitably evolves
into the absorbing state, and hence, the inactive phase is not qualitatively
affected by the temporal disorder. Naturally, the decay dynamics change
whether $\lambda=\lambda_{+}$ or $\lambda_{-}$.

Likewise, the active phase is also unaffected by disorder ($\lambda^{*}<\lambda_{\pm}\leq1$).
Evidently, the steady-state density $\rho_{\infty}$ fluctuates between
the corresponding values $\rho_{\infty}^{(S)}(\lambda_{-})$ and $\rho_{\infty}^{(S)}(\lambda_{+})$,
but the main feature of supporting long-standing activity regardless
of the initial state (provided that $\rho_{0}\neq0$) is unaffected.

The analysis of the metastable phase is more involving. Since $\rho_{\infty}^{(U)}(\lambda_{+})<\rho_{\infty}^{(U)}(\lambda_{-})$
{[}see, e.g., Eq.~\eqref{eq:r-infinity-QCP} and Fig.~\ref{fig:rho-lambda-scpmodel}{]},
when the initial state density $\rho_{0}\geq\rho_{\infty}^{(U)}(\lambda_{-})$
{[}$\rho_{0}\leq\rho_{\infty}^{(U)}(\lambda_{+})${]}, the system
will evolve to the active {[}inactive{]} state just like in the clean
metastable phase. The new feature happens when $\rho_{\infty}^{(U)}(\lambda_{+})<\rho_{0}<\rho_{\infty}^{(U)}(\lambda_{-})$.
In this case, the fate of the density will depend on the details of
the temporal fluctuation. If a rare fluctuation of long activity window
appears in the beginning, i.e., if initially $\lambda=\lambda_{+}$
for a sufficiently long period, the density then increases beyond
$\rho_{\infty}^{(U)}(\lambda_{-})$ and the system will thus evolve
towards the long-standing activity. On the other hand if this rare
fluctuation is such that$\lambda=\lambda_{-}$, then $\rho$ will
become less than $\rho_{\infty}^{(U)}(\lambda_{+})$ putting the system
towards inactivity. The lack of determinism for the evolution of $\rho(t)$
based only on knowledge of the initial condition $\rho_{0}$ is a
new feature appearing in the metastable phase due to temporal disorder
in the region $\rho_{\infty}^{(U)}(\lambda_{+})<\rho_{0}<\rho_{\infty}^{(U)}(\lambda_{-})$.

\subsubsection{Effects on the phase transitions}

Let us now discuss the more interesting cases when $\lambda_{-}$
and $\lambda_{+}$ are in different phases of the clean phase diagram.
We start analyzing the case when there is a mix of the inactive ($\lambda_{-}<\lambda_{c}$)
and the metastable ($\lambda_{c}\leq\lambda_{+}<\lambda^{*}$) phases.
Here, temporal disorder destroys the metastable phase replacing it
by the inactive one. The explanation is simple. After a sufficiently
long time, the system encounters with probability one a rare fluctuation
in which $\lambda=\lambda_{-}$ for a sufficiently long time interval
{[}greater than $T$ in Eq.~\eqref{eq:T}{]}. When this happens,
$\rho$ evolves below $\rho_{\infty}^{(U)}(\lambda_{+})$ and thus,
the system activity decays towards extinction. In addition, notice
that the first-order character of the transition between the inactive
and metastable phase (happening when $\lambda_{-}\rightarrow\lambda_{c}$)
is preserved. 

Because extinction happens only after a large and rare temporal interval
in the inactive phase, we call this phase as temporal Griffiths inactive
phase. Evidently, confirming the complete destruction of the metastable
phase numerically is a difficult task since the time $T^{\prime}$
needed for $\rho$ evolving below $\rho_{\infty}^{(U)}(\lambda_{+})$
is exponentially large in the interesting regime of $\lambda_{-}$
being sufficiently close to $\lambda_{c}$ (or $\Delta t\ll T$) and
$\lambda_{+}$ being far from $\lambda_{c}$. On average, the upper
limit time for decaying into the absorbing state is given by (see
Appendix \eqref{sec:Decaying-time}) 
\begin{equation}
\ln\overline{T^{\prime}}\sim-\frac{\ln p}{\Delta t\left(\lambda_{c}-\lambda_{-}\right)^{\phi}},\label{eq:averageT}
\end{equation}
 with $p$ and $\Delta t$ defined in Eq.~\eqref{eq:binary-dist}
and the diverging $T\sim\left(\lambda_{c}-\lambda_{-}\right)^{-\phi}$,
as defined in Eq.~\eqref{eq:T}. The fact that $T^{\prime}$ is very
different from $T$ when approaching the transition reinforces our
definition of temporal Griffiths inactive phase. In the usual quenched
(spatially) disordered case, the inactive phase near the transition
is called Griffiths phase because of the slower decay into the absorbing
state due to the existence of rare and large regions locally in the
active phase. In our case, however, a rare fluctuation in the inactive
phase is required.

\begin{figure}[t]
\begin{centering}
\includegraphics[clip,width=0.8\columnwidth]{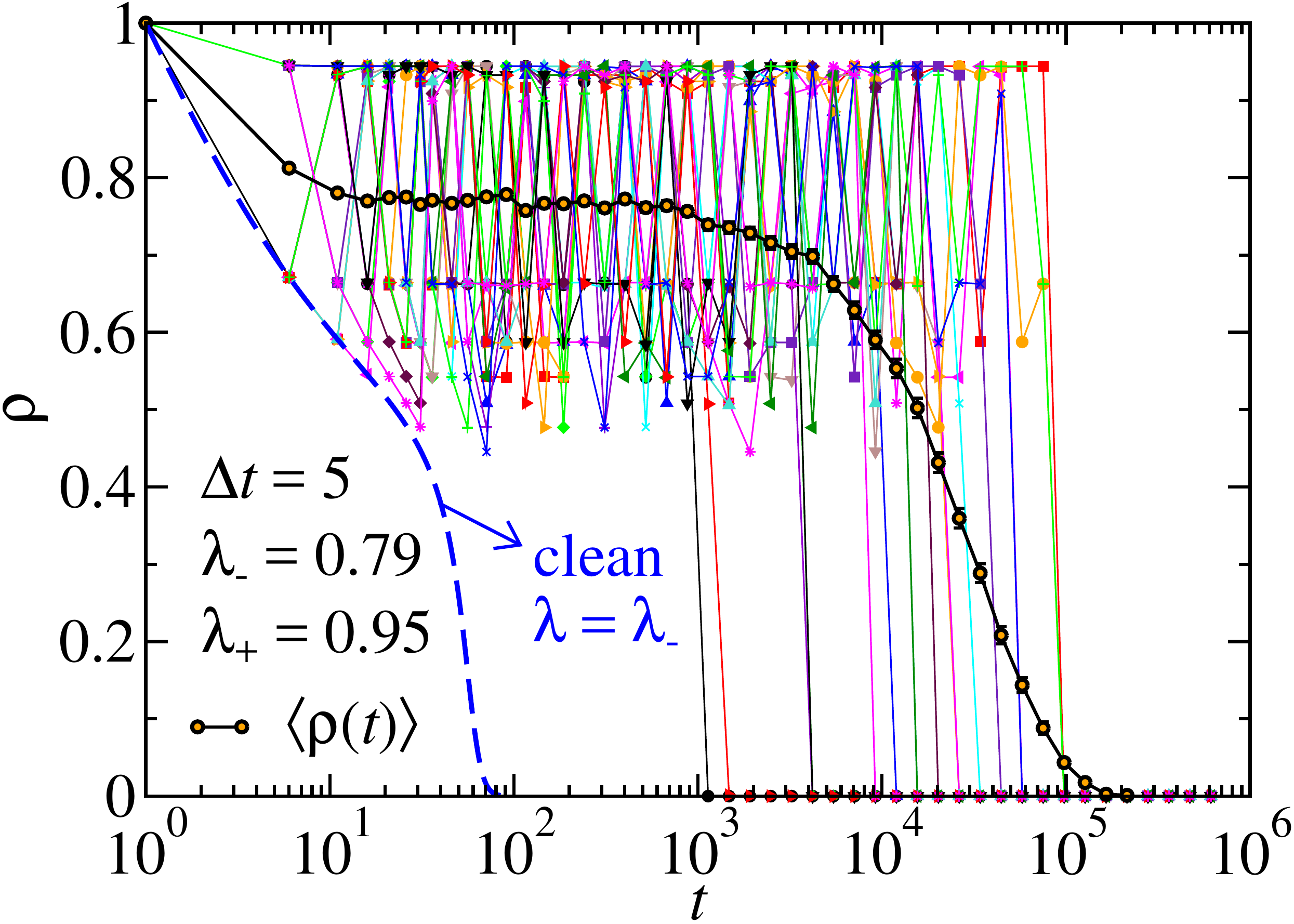}\\
\includegraphics[clip,width=0.8\columnwidth]{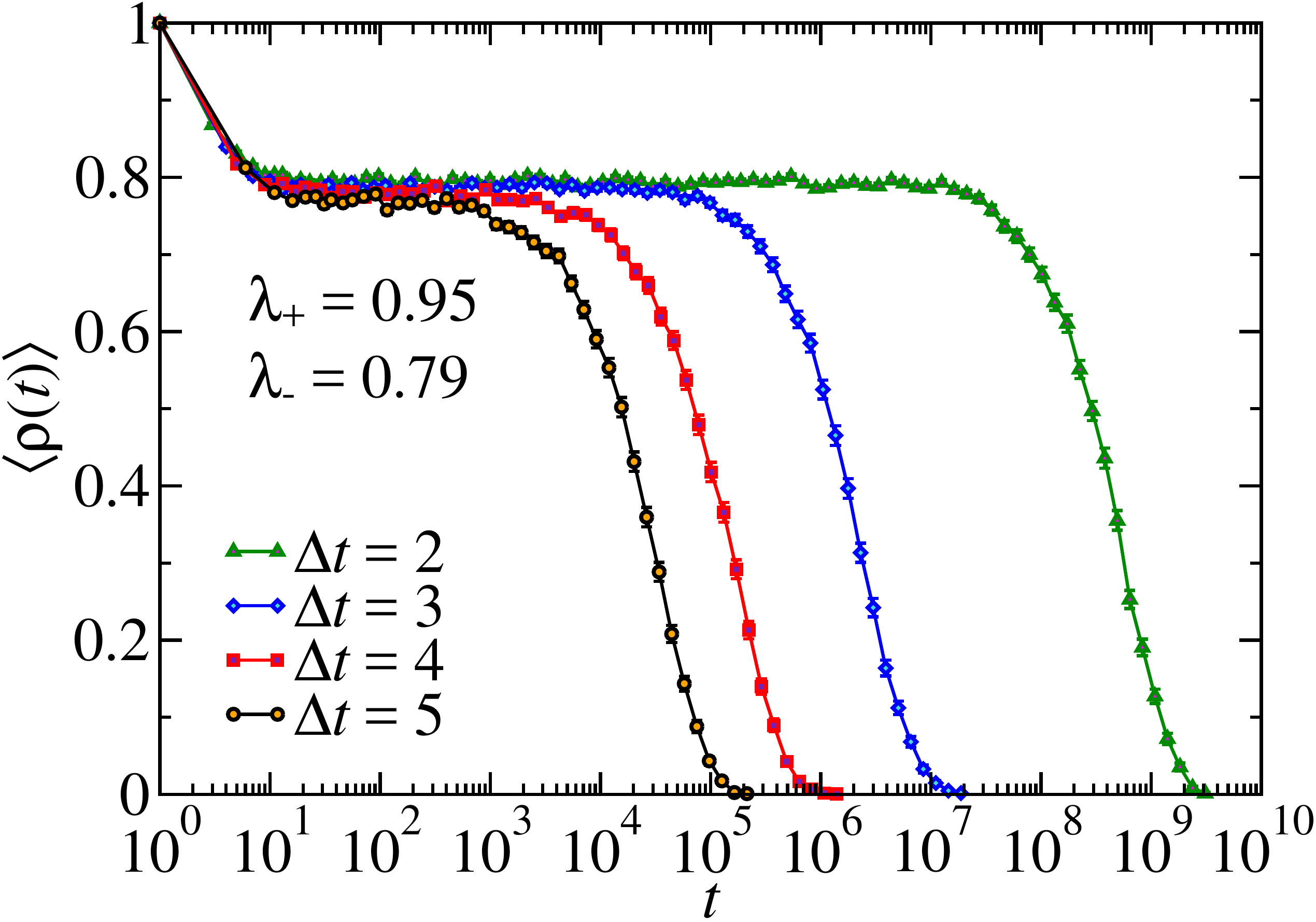}
\par\end{centering}

\caption{The mean-field density $\rho$ as a function of time for the QCP model.
The temporal disorder parameters are $p=\frac{1}{2}$, $\lambda_{-}=0.79$
and $\lambda_{+}=0.95$. In the top panel, $\rho$ is shown for $20$
disorder realizations by the data in various symbols and colors and
thin lines for $\Delta t=5$. The average density $\left\langle \rho\right\rangle $
(circles with thick lines) is obtained from $10^{3}$ disorder realizations.
In the bottom panel, the density is averaged for $10^{3}$ disorder
realizations for different time windows $\Delta t$. In all cases,
the lines connecting data symbols are guide to the eyes. \label{fig:rho-MF-QCP-random}}
\end{figure}

In order to illustrate the numerical effort for confirming the instability
of the Metastable phase towards the Temporal Griffiths Inactive one,
we plot in the top panel of Fig.~\eqref{fig:rho-MF-QCP-random} (for
clarity, only) $20$ different disorder realizations together with
the average over $10^{3}$ disorder realizations. Notice the large
spread of the decaying time for different samples, as a consequence,
the average $\rho$ decays smoothly over $2$ orders of magnitude.
Thus, we conclude that the average and typical decay times behave
very differently (another reason for associating this phase to Griffiths
physics). In addition, and most importantly for our discussion, notice
the difference between the decay times of the clean and random systems.
It rapidly increases for smaller time windows $\Delta t$ in accordance
with Eq.~\eqref{eq:averageT} as shown in the bottom panel. Even
though we have the analytical solution Eq.~\eqref{eq:rho-t-exact-QCP},
we could not reach the required time for the explicit demonstration
of the instability of the metastable phase for $\Delta t=1$, which
would happen for $T^{\prime}\sim10^{14}$. 

When $\lambda_{-}<\lambda_{c}$ (inactive phase) and $\lambda_{+}\geq\lambda^{*}$
(active phase), the actual system phase is decided by the analysis
of the low-density dynamics Eqs.~\eqref{eq:dr-dt-QCP} and \eqref{eq:dr-dt-sCP}. 

For the QCP model, the density decays exponentially in the inactive
phase as $\rho\sim e^{-(1-\lambda_{-})t}$ in the $\rho\rightarrow0$
limit. The active phase appears only when $\lambda=1$ and thus, $\partial_{t}\rho\sim\rho^{2}$.
Therefore $\rho$ grows much slower than the exponential. Consequently,
the system is in the temporal Griffiths inactive phase. 

For the $\sigma$CP model, on the other hand, the fate in the low-density
regime is determined by the competition between periods of inactivation,
in which $\rho\sim e^{^{2\left(\lambda_{-}-\lambda^{*}\right)t}}$,
with $\lambda^{*}=\frac{1}{2}$, and periods of activation, in which
$\rho\sim e^{^{2\left(\lambda_{+}-\lambda^{*}\right)t}}$. Therefore,
the system is in the active phase if $\lambda_{+}+\lambda_{-}>2\lambda^{*}$,
and it is in the inactive phase if $\overline{\lambda}<\lambda^{*}$.
For $\overline{\lambda}=\lambda^{*}$, the system is at the infinite-noise
critical point in the same universality class of the temporally disordered
CP model~\cite{vojta-hoyos-epl15}. Evidently, both the inactive
and active phases are of temporal Griffiths type. The latter has Griffiths
singularities in the same sense as in the contact process model with
temporal disorder in which the lifetime of finite systems does not
increase exponentially with the system volume (as in the pure active
phase) but rather as a power-law~\cite{vazquez-etal-prl11,vojta-hoyos-epl15,barghathi-etal-pre16}.

Finally let us analyze the case when there is a mix of the metastable
($\lambda_{c}\leq\lambda_{-}<\lambda^{*}$) and active ($\lambda_{+}\geq\lambda^{*}$)
phases. Again, we analyze details of the dynamics in the low-density
regime. Since the metastable phase behaves just as the inactive one
in the low-density regime, the same conclusions are  obtained for
$\lambda_{-}$ in the inactive and $\lambda_{+}$ in the active phases
applies.

\begin{figure}
\begin{centering}
\includegraphics[clip,width=0.8\columnwidth]{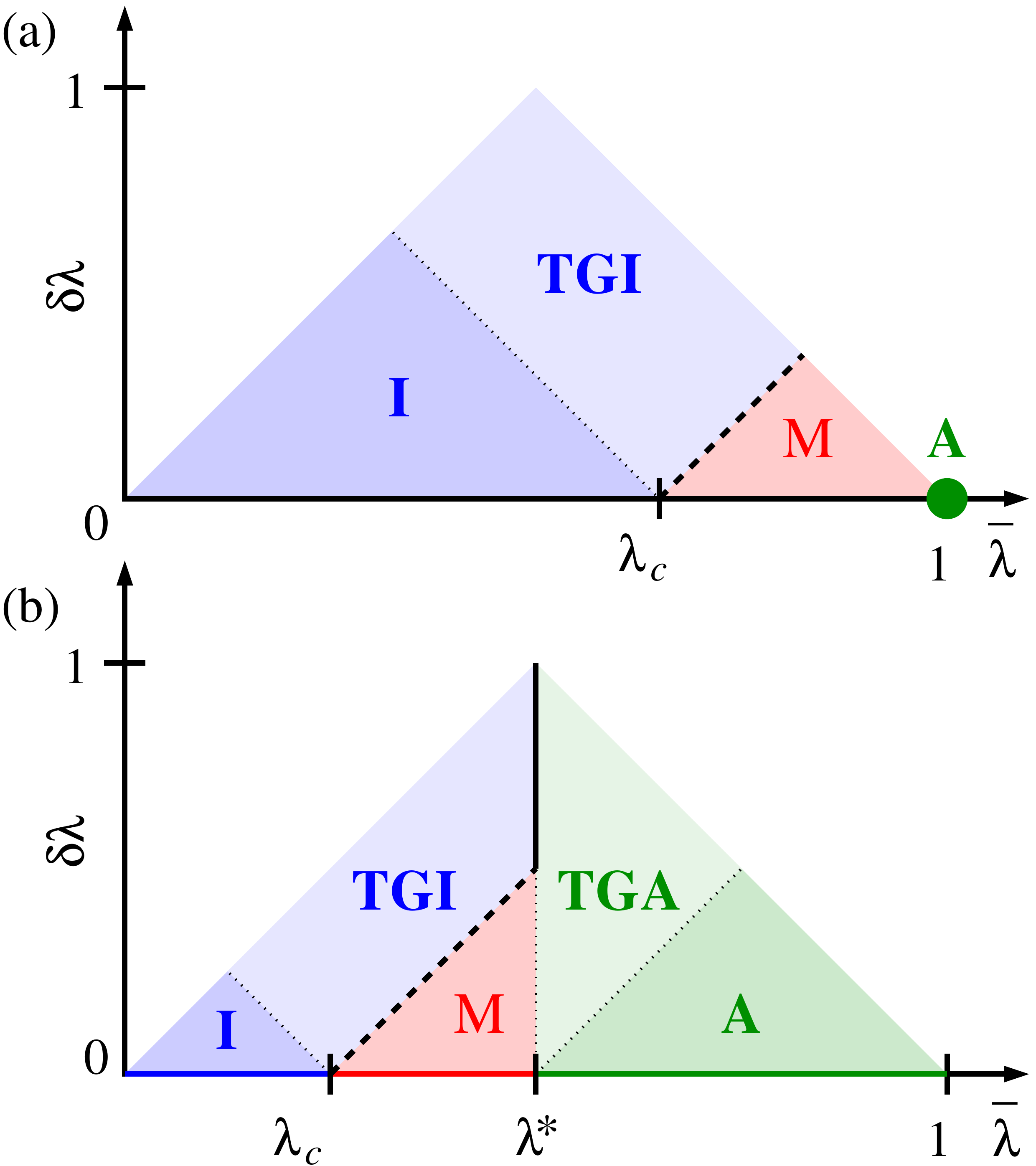} 
\par\end{centering}

\caption{The mean-field phase diagram in the temporally disordered case for
the (a) QCP and (b) $\sigma$CP models. The temporal disorder parameters
are defined in Eq.~\eqref{eq:binary-dist} with $p=1/2$. In the
$\sigma$CP model, we are considering that $a\zeta\left(\sigma\right)>1$,
otherwise the metastable phase vanishes. Dashed (solid) lines denote
first (second)-order phase transitions. Dotted lines represent crossovers.
The parameter space outside the shaded triangle is unphysical. For
the QCP model, $\lambda_{c}=\frac{4}{5}$ while it depends on $a$
and $\sigma$ for the $\sigma$CP model; and $\lambda^{*}=\frac{1}{2}$.
(A) stands for active, (I) for inactive, (M) for metastable and (TG)
for temporal Griffiths. \label{fig:dirty-PD}}
\end{figure}

We are now able to determine the mean-field phase diagram for the
QCP and $\sigma$CP models in the presence of temporal disorder as
shown in Fig.~\ref{fig:dirty-PD}. The dotted line are just crossovers.
The inactive and active phases, apart from trivial fluctuations, are
akin to the pure phases as discussed in Sec.~\ref{sub:Effects-on-phases}.
(Notice however that for the QCP model only the pure active phase
exists.) The temporally disordered metastable phase ($\delta\lambda\neq0$)
is also akin to the pure one except for the unpredictability of the
fate of the system state when the initial density is between the $\rho_{\infty}^{(U)}\left(\lambda_{+}\right)$
and $\rho_{\infty}^{(U)}\left(\lambda_{-}\right)$ as discussed in
Sec.~\ref{sub:Effects-on-phases}. The temporal Griffiths phases
have the same nature of their hosting phases but with different behaviors
due to rare temporal fluctuations. The dashed lines are metastable---inactive
first-order transitions while the solid line in the $\sigma$CP model
is a continuous inactive---active phase transition in the infinite-noise
universality class of the CP model. Finally, notice that this is the
first example of a non-equilibrium phase transition in which there
are temporal Griffiths phases in both sides of the transition.

\subsection{The probability density distribution for the density of active sites}

Due to noise (temporal disorder), the density of active sites greatly
fluctuates from sample to sample. It is thus desirable to obtain the
probability $R(\rho,t)$ of finding the system density between $\rho$
and $\rho+{\rm d}\rho$ at time $t$.

Let us start by analyzing the cases in which the density can become
arbitrarily small in the long-time limit. On then can be obtained
$R\left(\rho,t\right)$ using the methods of Ref.\ \onlinecite{vojta-hoyos-epl15},
where the logistic equations \eqref{eq:dr-dt-QCP} and \eqref{eq:dr-dt-sCP}
are linearized. In this approximation, the problem can be mapped into
a random walk problem for $x=-\ln\rho$. The nonlinear terms are then
replaced by a reflecting wall at the origin ensuring that the walker
position is always $x\geq0$ ($\rho\leq1$). Therefore, the probability
density distribution becomes 
\begin{equation}
Q\left(x,t\right)=\sqrt{\frac{2}{\pi\sigma_{v}^{2}n}}e^{-\frac{\left(x-\bar{v}n\right)^{2}}{2\sigma_{a}^{2}n}}-2\frac{\bar{v}}{\sigma_{v}^{2}}e^{\frac{2x\bar{v}}{\sigma_{v}^{2}}}\Phi\left(\frac{-x-vn}{\sigma_{v}\sqrt{n}}\right),\label{eq:Q(x,t)}
\end{equation}
 where $\Phi\left(z\right)=\frac{1}{\sqrt{2\pi}}\int_{-\infty}^{z}e^{-\frac{1}{2}y^{2}}{\rm d}y$
is the cumulative normal distribution, $\overline{v}$ and $\sigma_{v}$
are the random walker bias and bare width, respectively, and $n=t/\Delta t$
measures time in units of the time interval $\Delta t$. For the QCP
model, $\overline{v}=\overline{\mu\Delta t}=\left(1-\overline{\lambda}\right)\Delta t$
and $\sigma_{v}^{2}=\overline{\left(\mu\Delta t\right)^{2}}-\left(\overline{\mu\Delta t}\right)^{2}=\frac{1}{4}\delta\lambda^{2}\Delta t^{2}$,
whereas for the $\sigma$CP model, $\overline{v}=\overline{\left(\mu-\lambda\right)\Delta t}=2\left(\lambda^{*}-\overline{\lambda}\right)\Delta t$
and $\sigma_{v}^{2}=\delta\lambda^{2}\Delta t^{2}$.

The result \eqref{eq:Q(x,t)} is accurate far from the reflecting
wall and in the long time regime. Hence, in the inactive phase we
find that 
\begin{equation}
Q_{{\rm inactive}}(x,t)\approx\sqrt{\frac{\Delta t}{2\pi\sigma_{v}^{2}t}}e^{-\frac{(x-\overline{x}_{{\rm inactive}})^{2}}{2\sigma_{v}^{2}t/\Delta t}},\label{eq:Q-inactive}
\end{equation}
 where $\overline{x}_{{\rm inactive}}=\overline{v}n+\frac{\sigma_{v}^{2}}{2\bar{v}}+{\cal O}\left(t^{-1}\right)$
is the walker mean value, with the constant term being the leading
correction due to the reflecting wall. Notice that $Q_{{\rm inactive}}$
represents a simple random walker drifting away from the origin as
$t\rightarrow\infty$.

The result \eqref{eq:Q(x,t)} can also be applied to the active phase
close to the transition (which happens only for the $\sigma$CP model
for $\overline{\lambda}\apprge\lambda^{*}$), yielding 
\begin{equation}
Q_{{\rm active}}(x,t\rightarrow\infty)\approx-\frac{2\bar{v}}{\sigma_{v}^{2}}e^{\frac{2\bar{v}}{\sigma_{v}^{2}}x}=\frac{e^{-x/\overline{x}_{{\rm active}}}}{\overline{x}_{{\rm active}}},\label{eq:Q-active}
\end{equation}
where the walker mean value is $\overline{x}_{{\rm active}}=\frac{\sigma_{v}^{2}}{2\left|\overline{v}\right|}$.

Naturally, Eq.~\eqref{eq:Q(x,t)} also applies to the transition
between the active and inactive phases in which 
\begin{equation}
Q_{{\rm critical}}\left(x,t\right)\approx\sqrt{\frac{2\Delta t}{\pi\sigma_{v}^{2}t}}e^{-\frac{x^{2}\Delta t}{2\sigma_{v}^{2}t}}.\label{eq:Q-critical}
\end{equation}
Notice that $Q_{{\rm critical}}$ is a half Gaussian distribution
which broadens without limit as $t\rightarrow\infty$ illustrating
the infinite-noise criticality concept. This also implies that the
walker mean value is $\overline{x}_{{\rm critical}}=\sqrt{\frac{2\sigma_{v}^{2}t}{\pi\Delta t}}$.

The result \eqref{eq:Q(x,t)} can also be applied to the entire metastable
phase of both models if one starts with sufficiently small initial
densities {[}below $\rho_{\infty}^{(U)}(\lambda_{+})${]}. In this
case, the metastable phase behaves similarly to the inactive phase,
and hence, $Q_{{\rm inactive}}$ in Eq.~\eqref{eq:Q-inactive} accurately
describes the probability density distribution.

We now comment on the cases in which the density $\rho$ does not
become small. These happen for the active phase (of the $\sigma$CP
model) far away from the inactive phase and for the metastable phase
(of both models) provided that one starts with a sufficiently high
initial density {[}above $\rho_{\infty}^{(U)}(\lambda_{-})${]}. Clearly,
the density of active sites fluctuates between the values $\rho_{\infty}^{(S)}(\lambda_{-})$
and $\rho_{\infty}^{(S)}(\lambda_{+})$. Since the nonlinear terms
in Eqs.~\eqref{eq:dr-dt-QCP} and \eqref{eq:dr-dt-sCP} are important,
it becomes cumbersome to analytically predict the resulting stationary
probability density distribution $R(\rho,t)\rightarrow S(\rho)$.
For instance, if $\Delta t$ is much greater than the relaxation time
required to go from $\rho_{\infty}^{(S)}(\lambda_{-})$ to $\rho_{\infty}^{(S)}(\lambda_{+})$
(and vice-versa), then one mostly finds $\rho$ either very close
to $\rho_{\infty}^{(S)}(\lambda_{-})$ or $\rho_{\infty}^{(S)}(\lambda_{+})$.
Therefore, $S(\rho)$ is approximately a bimodal distribution peaked
around $\rho_{\infty}^{(S)}(\lambda_{-})$ and $\rho_{\infty}^{(S)}(\lambda_{+})$.
On the other hand for small $\Delta t$, the system has little time
to relax between $\rho_{\infty}^{(S)}(\lambda_{-})$ and $\rho_{\infty}^{(S)}(\lambda_{+})$.
Hence, $S(\rho)$ will be peaked at some value between $\rho_{\infty}^{(S)}(\lambda_{-})$
and $\rho_{\infty}^{(S)}(\lambda_{+})$.

Finally, let us analyze the last case in which both $\lambda_{+}$
and $\lambda_{-}$ are in the metastable phase and the initial density
is such that $\rho_{\infty}^{(U)}(\lambda_{+})<\rho_{0}<\rho_{\infty}^{(U)}(\lambda_{-})$.
As discussed in Sec.~\ref{sub:Overview}, the fate of the activity
depends on the details of the temporal disorder. If initially $\lambda=\lambda_{+}$
for a long interval of time, then the density will increase above
$\rho_{\infty}^{(U)}(\lambda_{-})$ and thus will remain finite in
the stationary regime. Otherwise if $\lambda=\lambda_{-}$ for a long
time window, the system then evolves towards the inactive absorbing
state. In this case therefore, the distribution of $\rho$ will have
two components resulting in $(1-\alpha)S(\rho)+\alpha\rho^{-1}Q_{{\rm inactive}}(e^{-\rho},t)$,
where $\alpha$ is the probability that the system evolves into the
absorbing state.

\begin{figure}[t]
\centering{}\includegraphics[width=0.9\columnwidth]{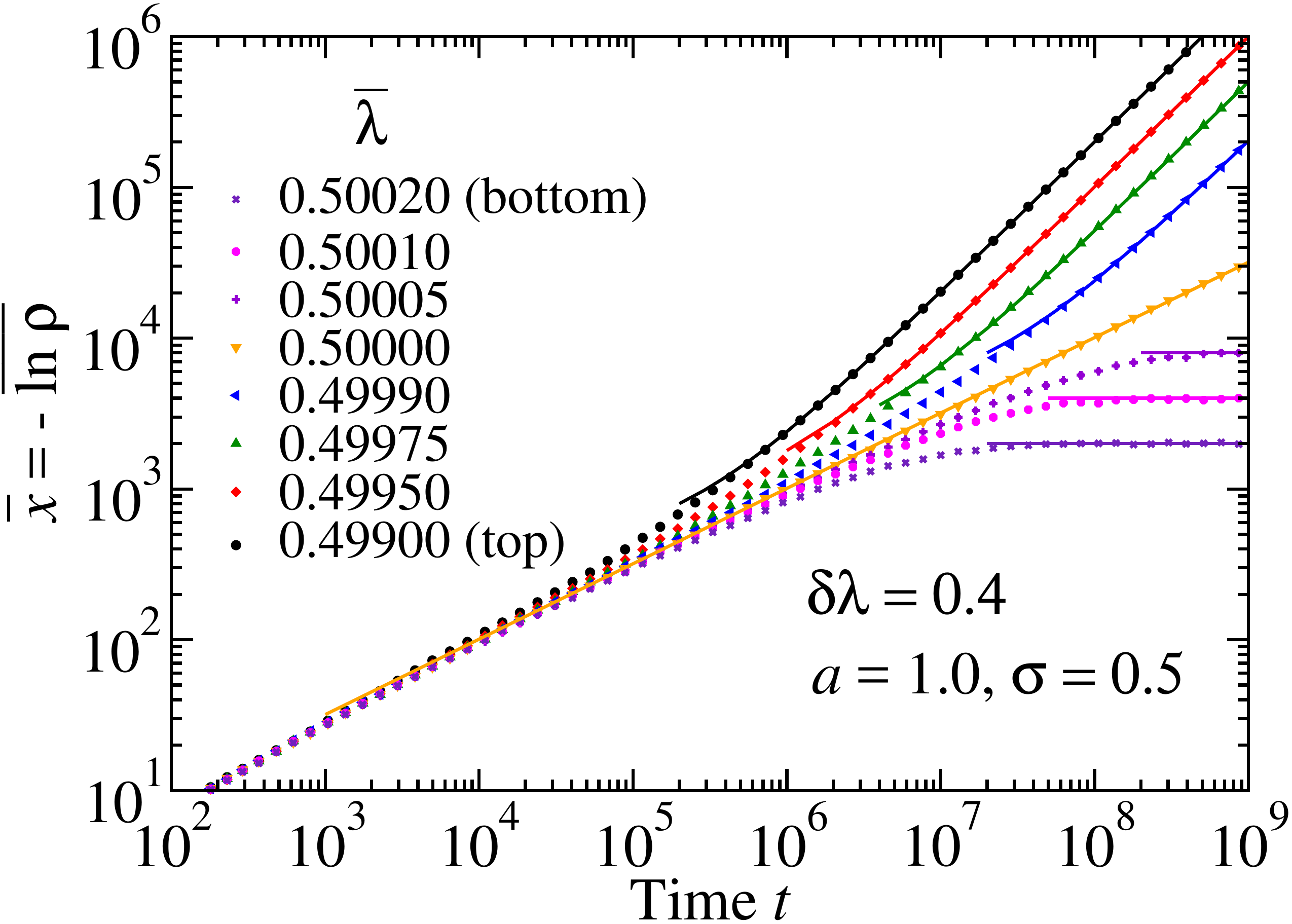}
\caption{The logarithm of the typical density as a function of the time $t$
for the $\sigma$CP model at the mean-field level. The long-range
interaction parameters are $a=1$ and $\sigma=0.5$. The temporal
disorder parameters are $\delta\lambda=0.4$ and $\Delta t=10$ {[}see
Eq.~(\ref{eq:binary-dist}){]}. Data points are averaged over $N=5\times10^{5}$
disorder realizations. Error bars are about the size of the symbols.
Solid lines are the analytical predictions (no fitting parameters)
based on the simple random walk picture (see main text).\label{fig:meanrho-t} }
\end{figure}

We report that we have confirmed all the above results by numerically
solving Eqs.~\eqref{eq:dr-dt-QCP} and \eqref{eq:dr-dt-sCP} in the
presence of temporal disorder via the Euler method and then computing
the corresponding probability density distribution. Here, we only
show in Fig.~\ref{fig:meanrho-t} the logarithm of the typical density
as a function of time for parameters near the transition between the
inactive and active phases for the $\sigma$CP model. We also show
as solid lines the analytical prediction for the infinite-noise criticality
in the long-time regime as discussed after Eqs.~\eqref{eq:Q-inactive}---\eqref{eq:Q-critical}.
The agreement is remarkable.

\section{Monte Carlo simulations\label{sec:Monte-Carlo-simulations}}

Our Monte Carlo simulations were performed in the lowest dimensions
in which both models exhibit a first-order phase transition: $d=2$
and $d=1$ for the QCP and the $\sigma$CP models, respectively~\cite{windus-jensen-jpa07,fiore-oliveira-pre07}.
In all cases, we consider periodic boundary conditions and $\mu=1-\lambda$
with $0<\lambda<1$. For the $\sigma$CP model, we have studied only
the case $\sigma=0.5$ and $a=2$.

As discussed in Sec.~\ref{sec:The-mean-field-approach}, simulations
of first-order transitions demands long computational times especially
in the presence of temporal disorder. For this reason, our purpose
is not to provide precise quantitative numbers, but rather confirm
the qualitative scenario of the temporal disorder effects on the first-order
phase transitions of Sec.~\ref{sec:The-mean-field-approach}. Hence,
we firstly review the clean system in order to confirm the metastability
of the active phase and the algebraically diverging time $T$ in Eq.~\eqref{eq:T}.
Then, we provide data supporting the instability of the metastable
phase towards the absorbing state when temporal disorder allows for
fluctuations into the inactive phase. Finally, we confirmed the infinite-noise
criticality governing the transition between the inactive and active
phases which takes place in the strong disorder regime of the $\sigma$CP
model. We emphasize that it is not our purpose to perform a careful
quantitative study. Thus, finite-size effects, unimportant for our
discussion, may be strongly present in our data.

\subsection{The Monte Carlo dynamics}

The actual dynamics is implemented following Ref.\ \onlinecite{dickman-pre99}.
In the 2D square lattice QCP model, an active site, say, $i$, is
randomly chosen among all $M$ active sites in the system. With probability
$\frac{\mu}{\mu+\lambda}=1-\lambda$, site $i$ becomes inactive whereas,
with complementary probability, one of its four nearest neighbor sites,
say, $j$, is randomly chosen. If $j$ is active, the system state
remains unchanged; if not, it will become active if there is at least
one pair of diagonal nearest-neighbors active sites. Otherwise, the
state remains unchanged. Finally, the time is increased by $1/M$.

The dynamics in the 1D $\sigma$CP model is very similar. After randomly
choosing a site $i$ among all the $M$ active ones, we also choose
with equal probability one of the two directions in the lattice. Then,
we compute the corresponding activity spreading rate $\lambda_{\ell}=\lambda\left(1+a\ell^{-\sigma}\right)$,
with $\ell$ being the distance (in units of lattice spacing) to the
next active site in the chosen direction. Afterwards, with probability
$\frac{\mu}{\mu+\lambda_{\ell}}=\frac{1-\lambda}{1+\lambda a\ell^{-\sigma}}$
the site $i$ becomes inactive whereas, with complementary probability,
the nearest-neighbor site in that chosen direction becomes active
(if it was already active, the system state remains unchanged). As
in the QCP model, the time is incremented by $1/M$. In these cases,
one performs averages over $N_{{\rm MC}}$ different Monte Carlo runs.
Since we also aim to study the metastable phase, we need as well to
perform simulations starting from a partially filled lattice in which
a fraction $0<\rho_{0}<1$ of sites (randomly chosen) is active.

Temporal disorder is implemented as explained in Sec.\ \ref{sec:The-models}.
We start with an activity spreading rate drawn from Eq.\ (\ref{eq:binary-dist}),
and whenever the many time increments sum $\Delta t$, a new $\lambda$
is drawn from the same binary distribution.

In the usual clean CP model, one usually performs simulations averaging
over $N_{\text{MC}}$ different Monte Carlo runs. In our study, we
also need to average over $N_{\text{D}}$ different disorder realizations
of the temporal sequence $\{\lambda_{1},\,\lambda_{2},\,\lambda_{3},\,\dots\}$.
We verified that our results have no dependence on $N_{\text{MC}}$
as long as $N_{\text{D}}\gg1$, i.e., it is sufficient using only
one Monte Carlo run $N_{{\rm MC}}=1$ for a given temporal sequence
$\{\lambda_{i}\}$ provided that the number of different disorder
realizations $N_{\text{D}}$ is sufficiently large. In addition, because
we want to study the metastable phase, we need as well to perform
simulations starting from a partially filled lattice in which a fraction
$0<\rho_{0}<1$ of (randomly chosen) sites is active. Therefore, we
also need to average over $N_{\text{S}}$ different initial states
for each sequence $\{\lambda_{i}\}$. We report that only one different
state $N_{\text{S}}=1$ for each temporal sequence is sufficient for
obtained unbiased and reliable data as long as the number of different
disorder realizations $N_{\text{D}}$ is large. For these reasons,
in what follows, we present our data average averaged over $N_{\text{D}}=N$
disorder realizations. This means that only one Monte Carlo run $N_{\text{MC}}=1$
for each of these sequences were performed. For the cases in which
$0<\rho_{0}<1$, this also means that $N$ different initial states
were considered in the simulation.

\subsection{The clean system }

Let us start by analyzing the clean case. The metastability of the
active phase near a first-order non-equilibrium phase transition into
an absorbing state has been reported in the literature in many different
situations~\cite{mikhailov-book90,meerson-sasorov-pre11}. It persists
in any spatial dimension supporting a first-order phase transition
and we have confirmed it in both studied models. 

\begin{figure}[b]
\centering{}\includegraphics[clip,width=0.65\columnwidth]{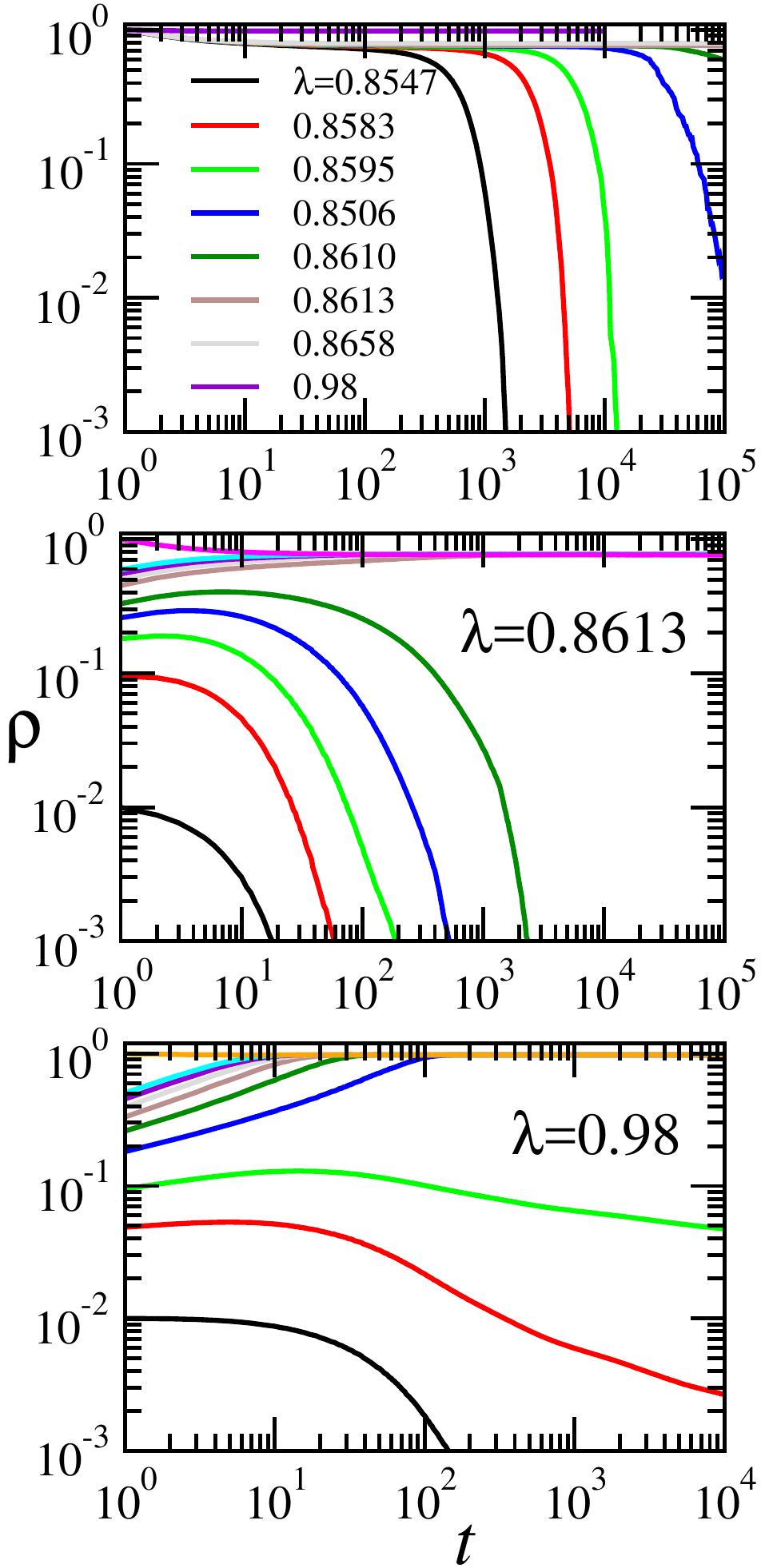} \caption{The average density as a function of the simulation time for the QCP.
In the top panel, $\rho$ is shown for various $\lambda$ starting
for $\rho_{0}=1$. The middle and bottom panels show $\rho$ for $\lambda=0.8613$
and $\lambda=0.98$ and various different initial densities $\rho_{0}$.
Data are averaged over $10^{2}$ (for the cases when $\rho$ is large
for large $t$)---$10^{5}$ (otherwise) different Monte Carlo runs
for systems of linear size $L=200$.\label{fig:QCP-clean} }
\end{figure}

In Fig.\ \ref{fig:QCP-clean}, we plot the average density of active
sites $\rho(t)$ as a function of time $t$ for the QCP model for
systems of linear size $L=200$ (we have also used $L=400$ and verified
the same conclusions). In the top panel, the initial density is fixed
at $\rho_{0}=1$ and the activity spreading rate $\lambda$ is varied.
In the remaining panels,$\rho_{0}$ is varied while $\lambda$ is
fixed at $0.8613$ (middle) and $0.98$ (bottom). From the top and
middle panels we conclude that a first-order phase transition takes
place at $0.8610<\lambda_{c}\leq0.8613$. Interestingly, we conclude
from the bottom panel that, in similarity with the mean-field results
of Sec.\ \ref{sub:Mean-field-QCP}, the active phase of the QCP model
is entirely metastable (except for the trivial case $\lambda=1$).
We have also confirmed it for slightly different implementations of
the dynamics and for $\lambda=0.99$. We thus conjecture that this
is a general feature of the active phase of the QCP model for any
spatial dimension $d\geq2$. 

\begin{figure}[t]
\centering{}\includegraphics[clip,width=0.65\columnwidth]{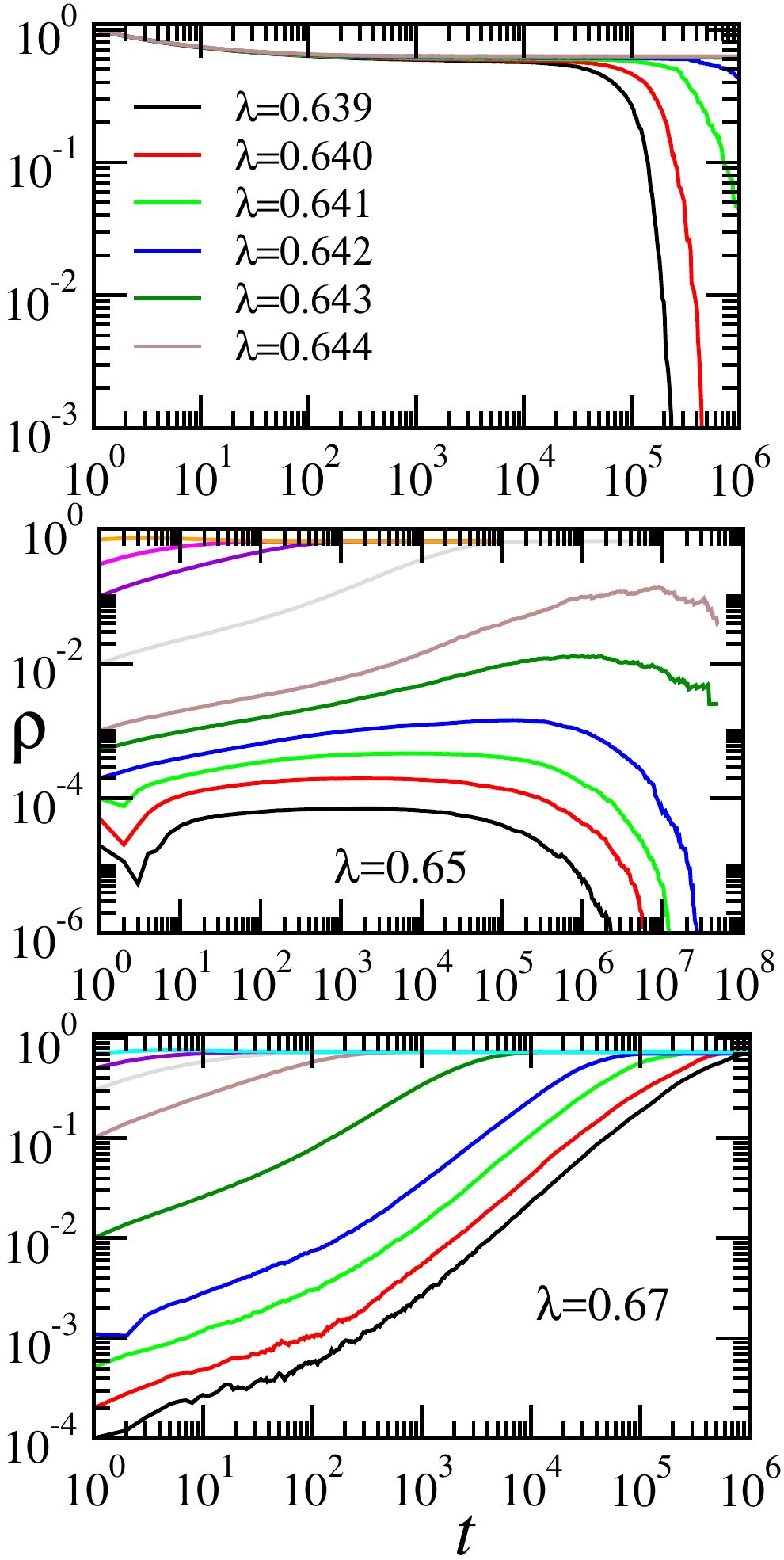} \caption{Similar to Fig.~\ref{fig:QCP-clean} but for the $\sigma$CP model
with $a=2.0$, $\sigma=0.5$, $L=10^{5}$ and the data are averaged
over $10^{3}$---$10^{5}$ different Monte Carlo runs. \label{fig:sCP-clean}}
\end{figure}

In Fig.~\ref{fig:sCP-clean} we study the $\sigma$CP model for $a=2.0$
and $\sigma=0.5$ for systems size $L=10^{5}$. As in the mean-field
approach, we find an active metastable phase in the interval $\lambda_{c}\leq\lambda<\lambda^{*}$
where we have identified $\lambda_{c}\ge0.643$ and $\lambda^{*}\leq0.670$. 

\begin{figure}[t]
\centering{}\includegraphics[width=0.8\columnwidth]{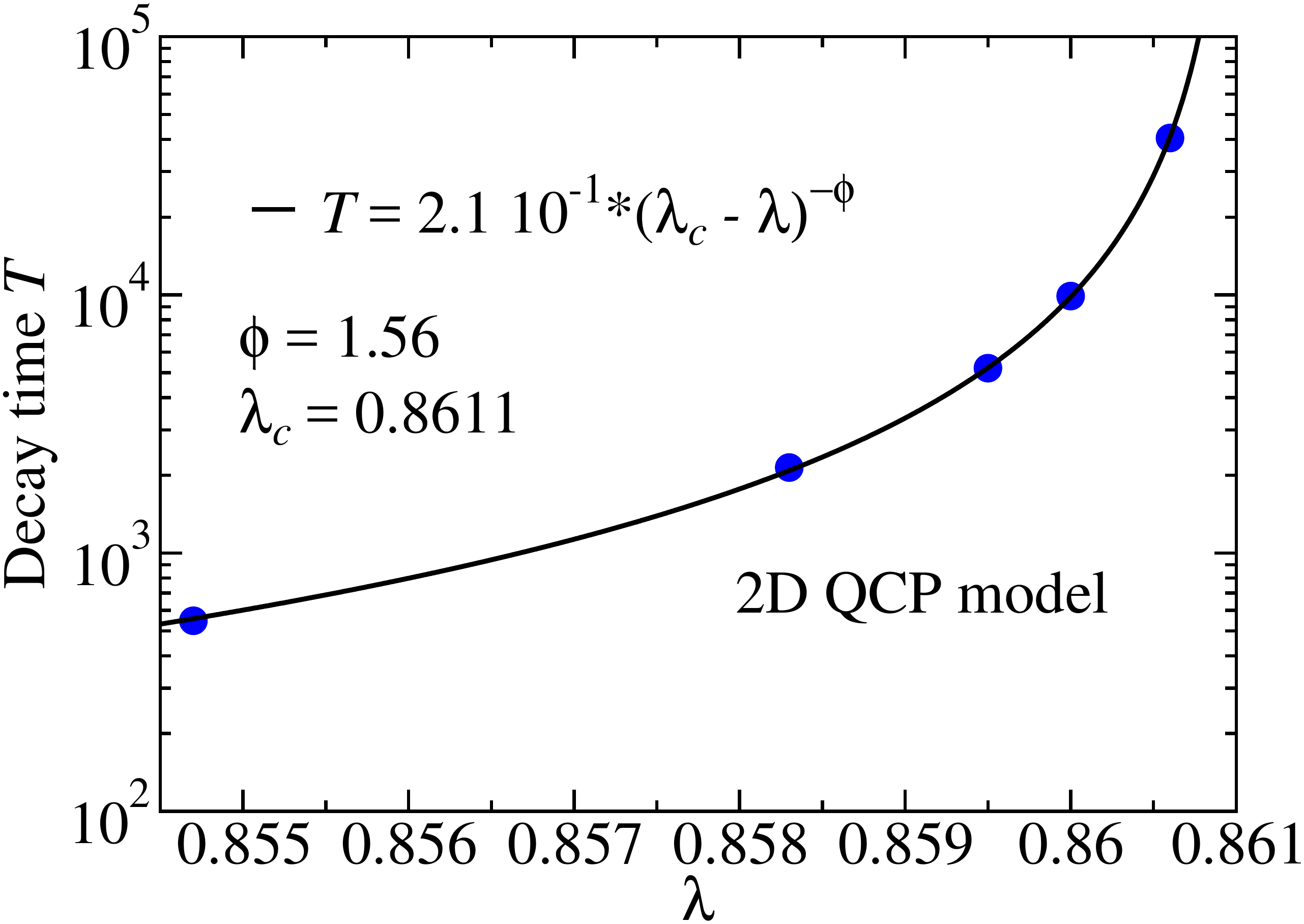}\\
\includegraphics[width=0.8\columnwidth]{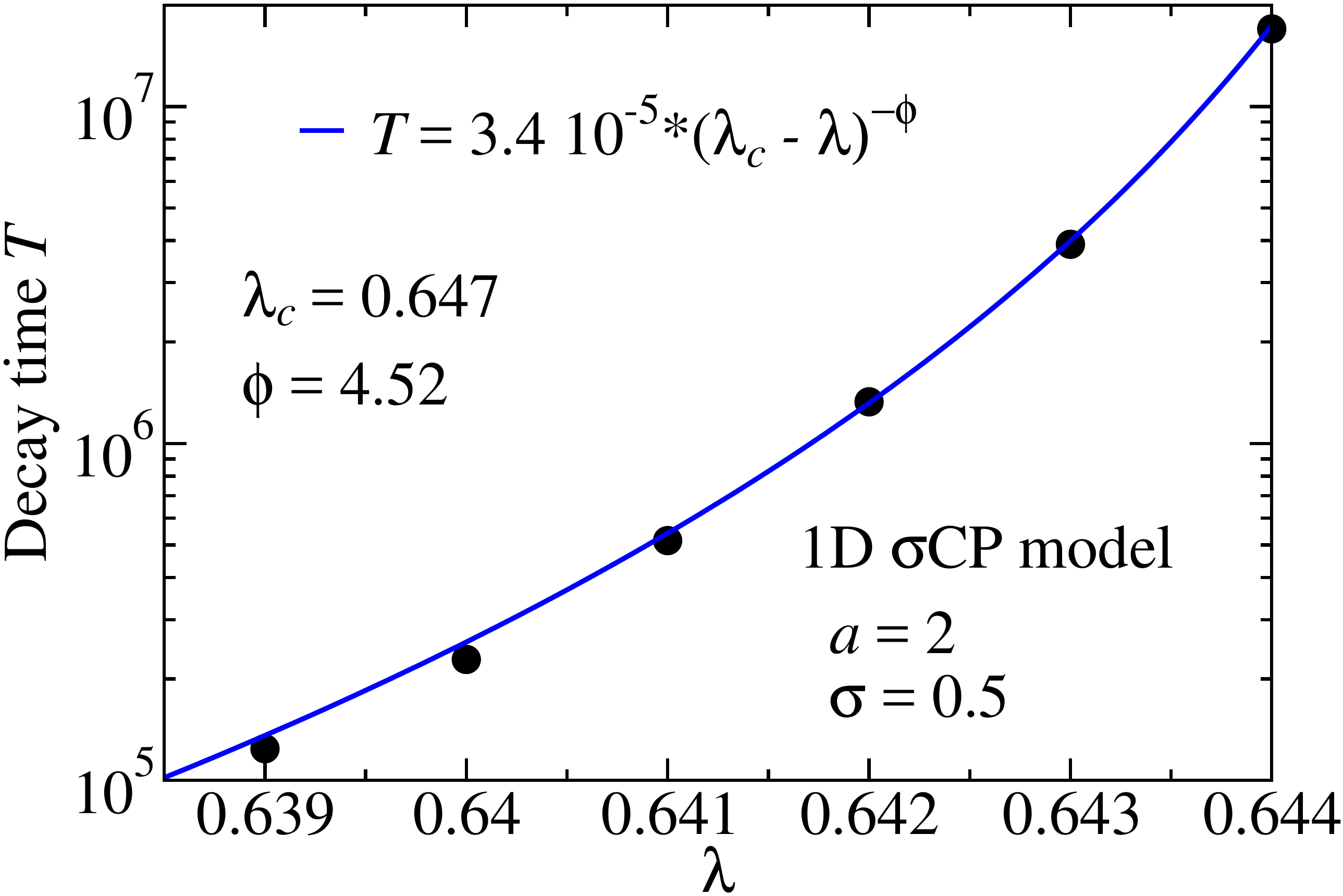} \caption{The decay time $T$ to the system toward the absorbing state as a
function of s $\lambda$ for the (top) QCP and (bottom) $\sigma$CP
models. In both cases, we observe an algebraic behavior of type $(\lambda_{c}-\lambda)^{-\phi}$,
where $\phi=1.56(1)$ and $4.52(3)$, respectively. \label{fig:decay-time}}
\end{figure}

We close this section by studying the time $T\sim\left(\lambda_{c}-\lambda\right)^{-\phi}$
required for the system decaying into the absorbing state as $\lambda\rightarrow\lambda_{c}$
(see Fig.~\ref{fig:decay-time}). We estimate $T$ from the data
on the top panels of Figs.~\ref{fig:QCP-clean} and \ref{fig:sCP-clean}
when $\rho(T)=0.1$. (We have used further data with fewer statistics
which are not shown.) We find the decay exponent $\phi\approx1.56(1)$
and $4.52(1)$ for the QCP and $\sigma$CP, respectively. Also, we
obtain the transition points $\lambda_{c}=0.8611(3)$ and $0.647(2)$
(for the QCP and $\sigma$CP models, respectively) from the data fitting.
Notice that we could not study $T$ for more than 2 orders of magnitude
close to the transition point and thus, our estimate may be plagued
with large systematic errors. 

\subsection{Temporal disorder}

\begin{figure}[t]
\centering{}\includegraphics[clip,width=1\columnwidth]{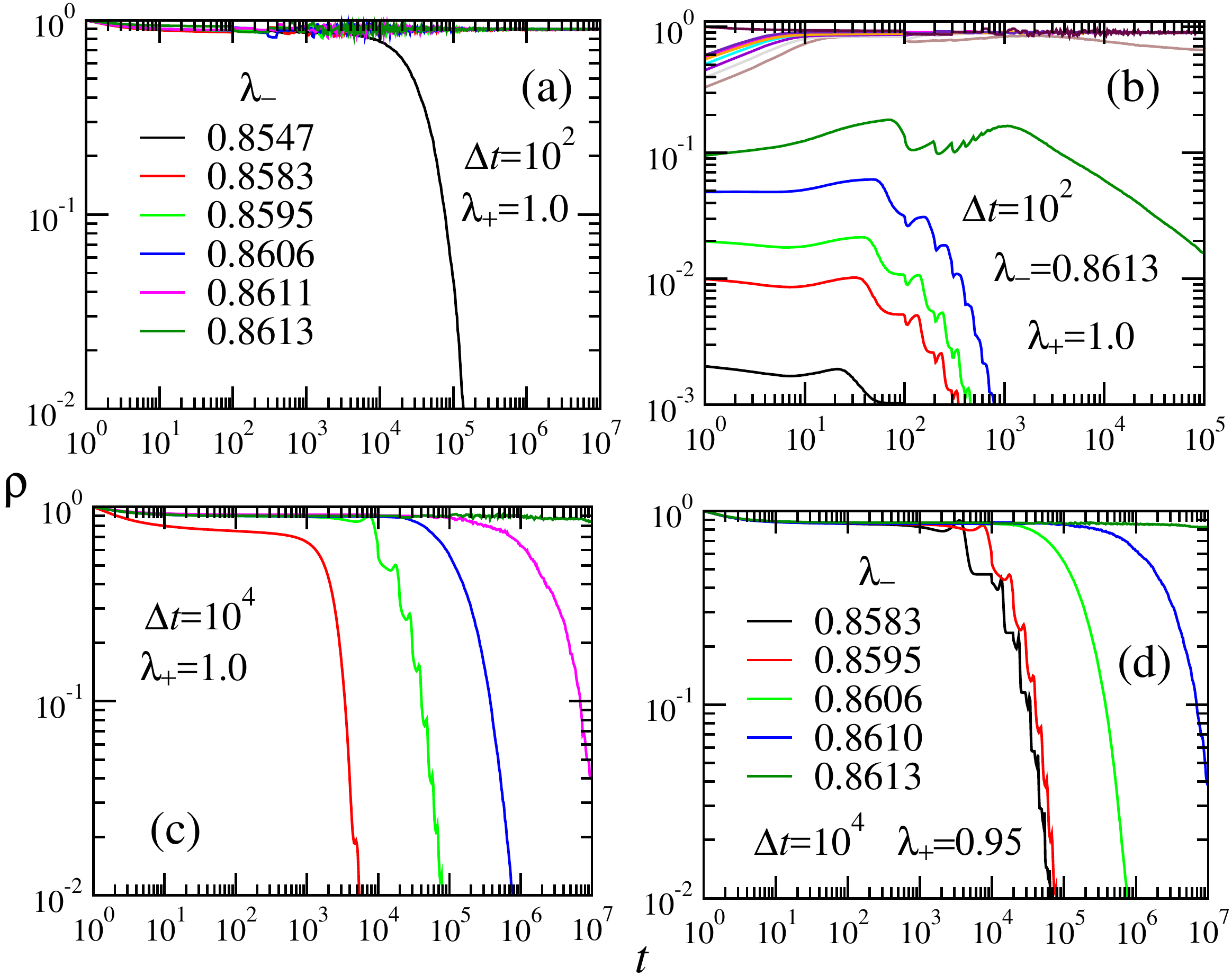}
\caption{The average density $\rho$ as a function of time $t$ for the 2D
QCP model for systems of size $L=200$ averaged over $N_{\text{MC}}=10^{2}$---$10^{5}$
disorder realizations. The disorder parameters {[}see Eq.~\eqref{eq:binary-dist}{]}
$\lambda_{\pm}$ are indicated in the legends {[}the one in panel
(a) also applies to (c){]}, $p=\frac{1}{2}$, and $\Delta t=10^{2}$
for panels (a) and (b) and $\Delta t=10^{4}$ for (c) and (d). In
all panels the initial density is $\rho_{0}=1$ except in panel (b)
where $\rho_{0}$ is varied.\label{fig:QCP-dirty} }
\end{figure}

We start our study analyzing the 2D QCP model. In panels (a) and (c)
of Fig.~\ref{fig:QCP-dirty} we confirm the instability of the active
phase ($\lambda_{+}=1.0$) with respect to temporal fluctuations into
the inactive phase $\lambda_{-}<\lambda_{c}\approx0.8613$. In panel
(a), we show for various $\lambda_{-}$ close to $\lambda_{c}$ that
the systems do not decay into the absorbing state up to large times
$\sim10^{7}$, which could be naively interpreted as the system being
active. However, as discussed in Eq.~\eqref{eq:averageT}, this is
not the case because the required time for decaying is extremely large.
Increasing $\Delta t$ to $10^{4}$ {[}see panel (c){]} reveals the
instability of the active phase of the 2D QCP model, just as in the
mean-field approach. Panel (b) of Fig.~\eqref{fig:QCP-dirty} corroborates
the metastability of the transition point $\lambda_{-}=\lambda_{c}$
and $\lambda_{+}=1$ between the inactive to the active phase of the
QCP model, and therefore, confirms the preservation of the first-order
transition character with respect to temporal disorder in the QCP
model. Finally, similar to panel (c), in panel (d) we confirm the
instability of the metastable phase ($\lambda_{c}<\lambda_{+}=0.95<\lambda^{*}=1$)
towards the absorbing state. We report that the transition point $\lambda_{-}=\lambda_{c}$
and $\lambda_{+}=0.95$ is also metastable {[}as in panel (b){]}.
Finally,we conclude that the phase diagram for the random 2D QCP model
is just as the mean-field one shown in Fig.~\hyperref[fig:dirty-PD]{\ref{fig:dirty-PD}(a)}
with $\lambda_{c}\approx0.8613$.

\begin{figure}[t]
\centering{}\includegraphics[clip,width=1\columnwidth]{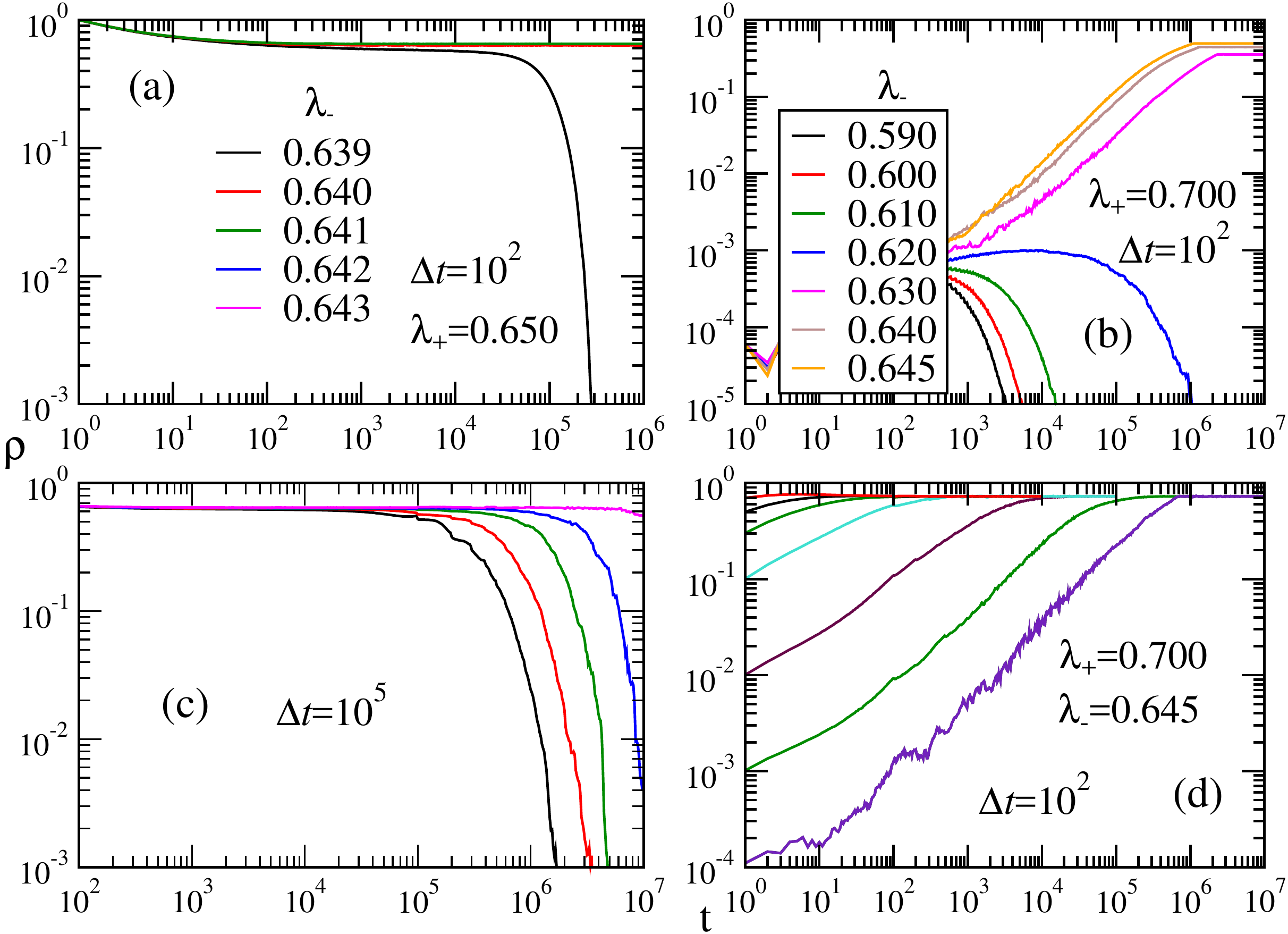}
\caption{The average density as a function of time for the disordered 1D $\sigma$CP
with disorder parameters {[}see Eq.~\eqref{eq:binary-dist}{]} $p=\frac{1}{2}$,
$\Delta t$ and $\lambda_{\pm}$ are indicated by the legends {[}with
the ones in panel (a) applying for panel (c) as well{]}. The system
size is $L=10^{5}$ averaged over $10^{2}$---$10^{4}$ disorder realizations.
In panel (a) and (c), $\lambda_{+}=0.65$ is in the metastable phase
while the various $\lambda_{-}\leq\lambda_{c}\approx0.643$ are in
the inactive phase. Panel (b) shows $\rho$ (starting from $\rho_{0}=5\times10^{-5}$)
for various $\lambda_{-}$ in the inactive phase while $\lambda_{+}=0.700$
is in the active one. Panel (d) $\rho$ starting from various different
initial conditions for $\lambda_{+}$ in the active phase while $\lambda$
is in the metastable one.\label{fig:sCP-dirty} }
\end{figure}

Figure \ref{fig:sCP-dirty} shows the main numerical results for the
1D $\sigma$CP. In panel (a) and (c) we plot $\rho(t)$ for many cases
in which $\lambda_{-}<\lambda_{c}\approx0.643$ is in the inactive
phase while $\lambda_{+}=0.650<\lambda^{*}\approx0.670$ is in the
metastable phase. As in the QCP model, the instability of the metastable
phase is manifest for the time window studied only when we consider
sufficiently large $\Delta t=10^{5}$ as shown in panel (c). Panel
(c) is analogous to panel (a) but $\lambda_{+}=0.700>\lambda^{*}$
is in the active phase, and the simulations start from $\rho_{0}=5\times10^{-5}$.
As can be seen, the active phase is stable for $\lambda_{-}\gtrapprox0.620$.
Finally, panel (d) shows $\rho(t)$ starting from different initial
conditions for $\lambda_{+}=0.700$ in the active phase and $\lambda_{-}=0.645$
in the metastable one. In this condition, it is clear that the system
is effectively active with no indications of bistability. Due to the
small range of the metastable phase ($0.643<\lambda<0.670$) we could
not reliably study the crossover line between the metastable and active
phases analogous to the dotted line in Fig.~\hyperref[fig:dirty-PD]{\ref{fig:dirty-PD}(b)}.

\begin{figure}[b]
\centering{}\includegraphics[clip,width=0.8\columnwidth]{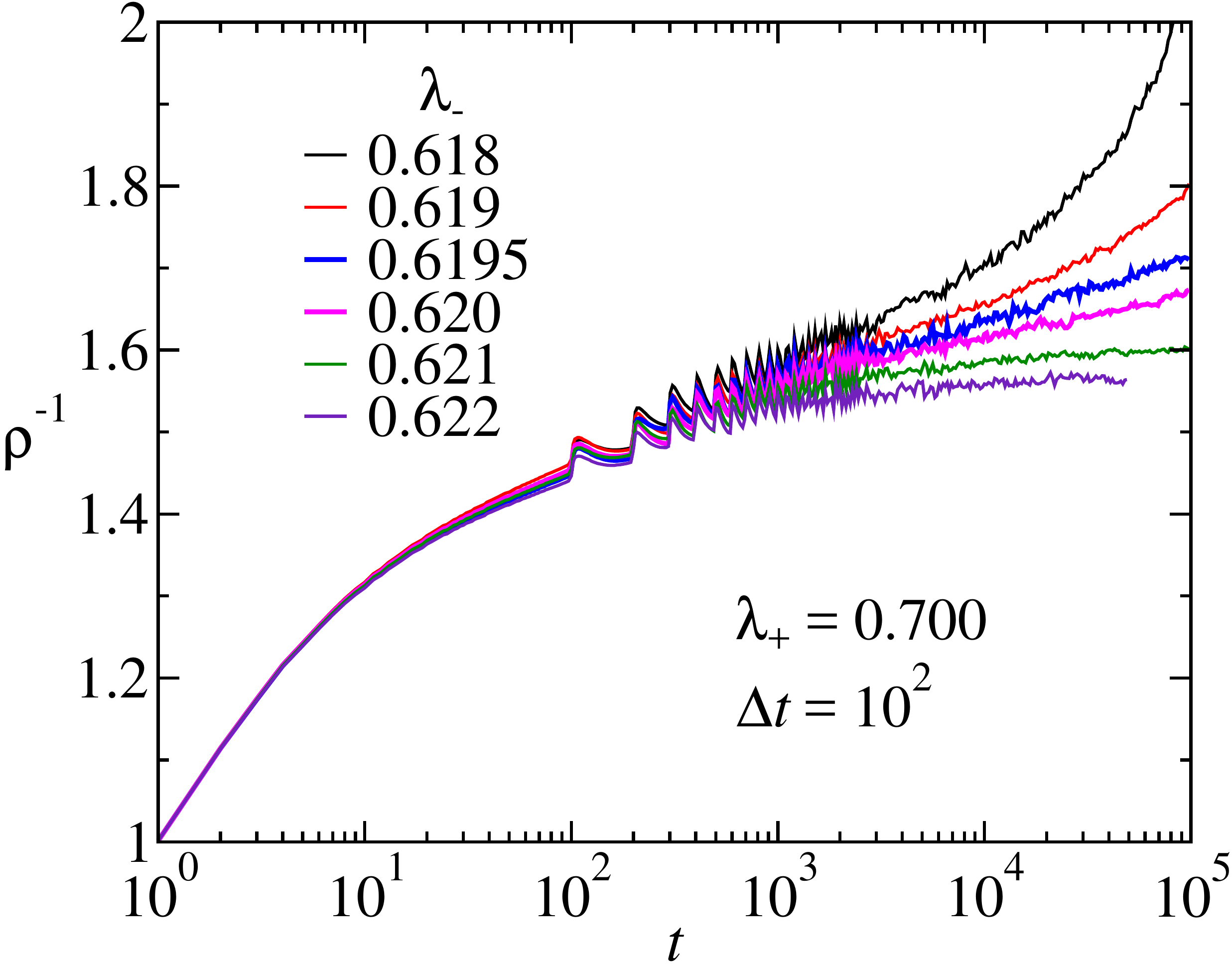}
\caption{The average density as a function of time for the $\sigma$CP with
$\lambda_{+}=0.7$, $p=\frac{1}{2}$,$\Delta t=10^{2}$ and various
distinct values of $\lambda_{-}$. The straight (blue) line has slope
$\rho\sim(\ln t)^{-1}$ for more than two orders of magnitude in $t$.
The system size is $L=10^{5}$ averaged over $10^{3}$---$3\times10^{3}$
different disorder realizations. \label{fig10} }
\end{figure}

As shown in Fig.~\hyperref[fig:dirty-PD]{\ref{fig:sCP-dirty}(b)},
there is a transition between the active and inactive phases for large
$\delta\lambda$. Our final numerical study is to confirm that this
transition is in the infinite-noise criticality. We then repeat the
study of Fig.~\hyperref[fig:dirty-PD]{\ref{fig:sCP-dirty}(b)} but
starting from the full lattice $\rho_{0}=1$ as shown in Fig.~\ref{fig10}.
We find that for $\lambda_{-}\approx0.6195(5)$ the system is critical
with average density vanishing (for 2 orders of magnitude in time)
as $\rho(t)\sim(\ln t)^{-1}$ ,exactly the same behavior of a system
in the infinite-noise criticality of the CP model~\cite{vojta-hoyos-epl15,barghathi-etal-pre16}.

Finally, we comment on the phase diagram of the random $\sigma$CP.
As in Fig.~\hyperref[fig:dirty-PD]{\ref{fig:dirty-PD}(b)}, the dashed
line representing the first-order phase transition is preserved in
any dimensions, i.e., its slope independs on $d$. For the studied
case ($a=2$ and $\sigma=0.5$ in $d=1$) we find that $\lambda_{c}\approx0.643$
and $\lambda^{*}\approx0.670$. We could not determine the dotted
line separating the bistability region from the usual active one.
For the continuous transition between the inactive and active phase
(solid line), we report that we have numerically verified that it
tilts to the right favoring the inactive phase. This is expected because
inactivation always provide an exponential decay of $\rho$ for any
dimension. On the other hand, only in the mean-field approximation
the activity spreads exponential fast. For finite dimensions, it can
only spread ballistically. Therefore, we expect a smaller active phases
when compared with mean field, and thus, the solid line must tilt
to the right.

\section{Conclusions\label{sec:disc}}

We have established a general theory of the effect of temporal disorder
in discontinuous non-equilibrium phase transitions into an absorbing
state. A quantitative analysis is present in the framework of mean-field
approach as well as numerical simulations in finite dimensions for
two paradigmatic models exhibiting first-order phase transitions,
namely the quadratic contact process (QCP) and the contact process
with long-range interactions ($\sigma$CP) models. Our work provides
an analytical basis for the numerical findings of Ref.~\onlinecite{oliveira-fiore-pre16}
that, in contrast to the spatial disorder, temporal disorder does
not forbid discontinuous transition in low dimensional systems. This
is not to be mistaken as a weaker effect in comparison since the metastable
active phase is replaced by the temporal Griffiths inactive phase. 

We have found that temporal disorder noise does not qualitatively
affect the phases when the fluctuations are confined within the phases,
except for small details in the metastable phase as discussed in Sec.~\ref{sub:Effects-on-phases}. 

On the other hand, the metastable phase is always unstable against
temporal disorder whenever it allows for fluctuations into the inactive
phase. Due to rare temporal fluctuations, the metastable phase becomes
a temporal Griffiths inactive phase in which the decay time become
exponentially large {[}see Eq.~\eqref{eq:averageT}{]}. Furthermore,
our general mean-field results show that the temporal Griffiths inactive
phase is a more general phenomena expected to appear in any non-equilibrium
first-order phase transition into an absorbing state.

For the QCP model, the active phase is also unstable against temporal
disorder and thus, only exists in the clean limit. As a consequence,
the first-order character of the transition is not destabilized by
temporal disorder for any disorder strength. In contrast for the $\sigma$CP
model, the active phase is robust against small fluctuations into
the inactive phase. As a consequence, the first-order transitions
is turned into a continuous one when the disorder strength is sufficiently
strong. In addition, we have found that the critical behavior belongs
to the infinite-noise universality class of the contact process model,
but with two Griffiths phases surrounding it.

Finally, we notice that the inactive phase being characterized by
an absorbing state is not a necessary condition for our theory. The
bistability of the active phase is. Therefore, although we have focused
only on two models, we expect that our theory applies to other models
exhibiting discontinuous non-equilibrium phases transitions such as,
e.g., the ZGB model~\cite{ziff-gulari-barshad-prl86} and the majority-vote
with inertia model~\cite{chen-etal-pre17}.

\section*{ACKNOWLEDGEMENT}

We acknowledge the financial support from CNPq, FAPESP and Simons
Foundation. JAH is grateful for the hospitality of the Aspen Center
for Physics.

\appendix
%dummy comment inserted by tex2lyx to ensure that this paragraph is not empty

\section{Decaying time near the temporal Griffiths inactive---metastable phase
transition\label{sec:Decaying-time}}

We intend to estimate the average time $\overline{T^{\prime}}$ for
decaying when the system undergoes a first-order phase transition
from the inactive to the metastable phase in the presence of temporal
disorder.

For simplicity, consider the case of binary disorder as defined in
Eq.~\eqref{eq:binary-dist} where $\lambda_{-}<\lambda_{c}$ places
the system in the inactive phase and $\lambda_{c}<\lambda_{+}<\lambda^{*}$
places the system in the metastable one. In this case, notice there
are only three relevant time scales in the problem: the time interval
$\Delta t$, the decay time $\tau_{-}$ (related to $\lambda_{-}$)
and the relaxation time $\tau_{+}$ (related to $\lambda_{+}$). Precisely,
the second is defined as the time required for the system to evolve
from $\rho^{(S)}(\lambda_{+})$ to $\rho^{(U)}(\lambda_{+})$ when
$\lambda=\lambda_{-}$ while the latter is the other way round when
$\lambda=\lambda_{+}$. 

Let us consider the case when the disordered system is close to the
metastable phase, thus $\lambda_{c}\gg\lambda_{c}-\lambda_{-}>0$
{[}implying $\tau_{-}\sim T$ in Eq.~\eqref{eq:T}{]}. For further
simplicity, consider the case $\tau_{+}\ll\Delta t\ll\tau_{-}$ (which
could possibly be accomplished when $\lambda_{+}$ is deep in the
metastable phase: $\lambda^{*}\gg\lambda^{*}-\lambda_{+}>0$). With
those assumptions, the only way of decaying into the absorbing state
(starting from the initial condition $\rho_{0}=1$) is via a sufficiently
long and continuous sequence of $k=\tau_{-}/\Delta t$ inactive time
intervals ($\lambda=\lambda_{-}$) such that $\rho$ becomes less
than $\rho^{(U)}(\lambda_{+})$ afterwards. 

Since this long sequence is rare, the waiting time $T^{\prime}$ can
be extremely long as we show in the following. Consider intervals
of duration $\tau_{-}$ which appear with probability $p_{\tau}=p^{k}$,
with $p$ being the probability for an interval being of inactive
type {[}see Eq.~\eqref{eq:binary-dist}{]}. Then, starting from an
active state, the probability that the system decays just after the
$n$th of such time intervals is 
\begin{equation}
P_{n}=\left(1-p_{\tau}\right)^{n-1}p_{\tau}.\label{eq:Pn}
\end{equation}
 Thus, the average waiting time for decaying into the absorbing state
is 
\[
\overline{T^{\prime}}\approx\tau_{-}\sum_{n=1}^{\infty}nP_{n}=\tau_{-}p^{-\frac{\tau_{-}}{\Delta t}},
\]
 and we recall that $\tau_{-}\sim T$.

\bibliography{referencias}

\end{document}